\documentclass[11pt, reqno]{article}

\usepackage[hmargin=1.0in,vmargin=1.0in]{geometry}
\usepackage{amsthm}
\usepackage[english]{babel}
\usepackage{cite}
\usepackage{color}
\usepackage{amsmath,amsfonts,amssymb,graphicx}
\usepackage{microtype}
\usepackage{mathrsfs,url}
\usepackage{enumerate}
\usepackage{cmathml}
\usepackage[lined,linesnumbered,commentsnumbered]{algorithm2e}
\usepackage{caption}
\usepackage{thm-restate}

\date{}

\newtheorem{theorem}{Theorem}
\newtheorem{lemma}{Lemma} 
\newtheorem*{lemma*}{Lemma} 
\newtheorem*{proposition*}{Proposition} 
\newtheorem*{theorem*}{Theorem} 
\newtheorem{proposition}[theorem]{Proposition}
\newtheorem{corollary}{Corollary}
\newtheorem{definition}{Definition}
\theoremstyle{remark}

\newcommand{\ignore}[1]{}
\newcommand{\tup}[1]{\ensuremath{\bar{#1}}}

\newcommand{\express}[1]{\ensuremath{\langle #1\rangle}}
\newcommand{\R}{\mbox{$\mathbb R$}}
\newcommand{\Q}{\mbox{$\mathbb Q$}}
\newcommand{\Qp}{\mbox{$\mathbb Q_{> 0}$}}
\newcommand{\Qnn}{\mbox{$\mathbb Q_{\geq 0}$}}
\newcommand{\range}{\Q}
\renewcommand{\cal}[1]{\mathcal{#1}}

\newcommand{\imp}[0]{{\rm Imp}}

\newcommand{\omegaour}[0]{\hat{\omega}}
\newcommand{\sigmaour}[0]{\hat{\sigma}}
\newcommand{\rhoour}[0]{\hat{\rho}}
\DeclareMathOperator{\sign}{sgn}
\newcommand{\ch}[0]{A} 

\DeclareMathOperator{\supp}{supp}
\newcommand{\opm}[2]{{\cal O}^{(#2)}_{#1}}
\newcommand{\sym}[1]{\bar{#1}}
\DeclareMathOperator{\argmin}{argmin}

\DeclareMathOperator{\opt}{Opt}

\begin{document}

\title{The complexity of finite-valued CSPs\thanks{
An extended abstract of this work appeared in the \emph{Proceedings of the
45th ACM Symposium on the Theory of Computing (STOC)}, pp. 695-704, 2013~\cite{tz13:stoc}.
Johan Thapper was partially supported by the European Research Council under
the European Community's Seventh Framework Programme
(FP7/2007-2013 Grant Agreement no.\ 257039).
Stanislav \v{Z}ivn\'y was supported by a Royal Society University Research
Fellowship.}}

\author{
Johan Thapper\\
Universit\'e Paris-Est, Marne-la-Vall\'ee, France\\
\texttt{thapper@u-pem.fr}
\and
Stanislav \v{Z}ivn\'{y}\\
University of Oxford, UK\\
\texttt{standa.zivny@cs.ox.ac.uk}
}

\maketitle

\begin{abstract}
  We study the computational complexity of exact minimisation of 
  rational-valued discrete functions. 
  Let $\Gamma$ be a set of rational-valued functions on a fixed finite domain;
  such a set is called a \emph{finite-valued constraint language}. The
  valued constraint satisfaction problem, VCSP$(\Gamma)$, is the problem of
  minimising a function given as a sum of functions from $\Gamma$. 
  We establish a dichotomy theorem with respect to exact solvability for
  \emph{all} finite-valued constraint languages defined on domains of \emph{arbitrary}
  finite size.

  We show that every constraint language $\Gamma$ either admits a binary
  symmetric fractional polymorphism in which case the basic
  linear programming relaxation solves
  any instance of VCSP$(\Gamma)$ exactly, or $\Gamma$ satisfies a simple
  hardness condition that allows for a polynomial-time reduction from
  Max-Cut to VCSP$(\Gamma)$. 
\end{abstract}

\section{Introduction}

In this paper we study the following problem: what classes of 
discrete extensionally-represented 
functions can be minimised exactly in polynomial time?
Such problems can be readily described as (finite-)valued constraint
satisfaction problems. We provide a complete answer to this question for
rational-valued functions defined on arbitrary finite domains. 

The constraint satisfaction problem, or CSP for short, provides a common
framework for many theoretical and practical problems in computer science.
Problems that can be cast in the CSP framework have been studied in several
contexts of computer science including artificial
intelligence~\cite{Dechter03:processing}, database
theory~\cite{Kolaitis00:jcss}, and graph
theory~\cite{Hell:graphs,Hell08:survey}. A CSP instance can informally be
described as a set of variables to be assigned values from the domains of the
variables so that all constraints are satisfied~\cite{Montanari74:constraints}.
The CSP is NP-complete in general and thus we are interested in restrictions
which give rise to tractable classes of problems. Following
\cite{Feder98:monotone}, we restrict the constraint language, that is, all
constraint relations in a given instance must belong to a fixed, finite set of
relations on the domain. The most successful approach to classifying
language-restricted CSPs is the so-called algebraic
approach~\cite{Jeavons97:closure,Jeavons98:algebraic,Bulatov05:classifying},
which has led to several complexity
classifications~\cite{Bulatov06:3-elementJACM,Bulatov11:conservative,Barto09:siam,Barto11:lics}
and algorithmic characterisations~\cite{Barto14:jacm,Idziak10:siam,Berman10:few}
going beyond the seminal work of Schaefer on Boolean
CSPs~\cite{Schaefer78:complexity}.

Several natural optimisation variants of CSPs have been studied in the
literature such as Max-CSP, where the goal is to maximise the number of
satisfied constraints (or, equivalently, minimise the number of unsatisfied
constraints)~\cite{Cohen05:supermodular,Creignouetal:siam01,JonssonKK06:3valued,Jonsson11:cp,DeinekoJKK08},
and Max-Ones~\cite{Creignouetal:siam01,Jonsson08:siam} and
Min-Cost-Hom~\cite{Takhanov10:stacs,Takhanov10:cocoon,Uppman13:icalp,Uppman14:stacs}, where all
constraints have to be satisfied and some additional function of the assignment
is optimised. The most general variant is the valued constraint satisfaction
problem, or VCSP for short, which deals with both feasibility and optimisation~\cite{Cohen06:complexitysoft,z12:complexity}. A
valued constraint language $\Gamma$ is a set of functions on a fixed domain and
a VCSP instance over $\Gamma$ is given by a sum of functions from $\Gamma$ with
the goal to minimise the sum. The VCSP framework is very robust and has also
been studied under different names such as Min-Sum problems, Gibbs energy
minimisation, Markov Random Fields, Conditional Random Fields and others in
different contexts in computer
science~\cite{Lauritzen96,Wainwright08,Crama11:book}. The VCSP in its full
generality considers functions with the range being the rationals with positive
infinity~\cite{Cohen06:complexitysoft}; this includes both CSPs (feasibility)
and Max-CSPs (optimisation) as
special cases where the range of the functions is $\{0,\infty\}$ and $\{0,1\}$,
respectively. In this work we will focus on finite-valued VCSPs, that is, the
range of the functions is the set of rationals. Finite-valued CSPs capture
optimisation problems. (Finite-valued CSPs are called
generalised CSPs in the approximation community~\cite{Raghavendra08:stoc}.)

Given the generality of the VCSP, it is not surprising that only few complexity
classifications are known. In the general-valued case (that is, when the range
of the functions is the rationals with positive infinity), only constraint languages on a
two-element domain~\cite{Cohen06:complexitysoft,cz11:cp-mwc} and conservative
(containing all $\{0,1\}$-valued unary functions) constraint languages~\cite{kz13:jacm}
have been completely classified with respect to exact solvability. In the
finite-valued case, constraint languages on two-element
domains~\cite{Cohen06:complexitysoft}, three-element domains~\cite{hkp14:sicomp},
and conservative constraint languages~\cite{kz13:jacm} have been completely classified with
respect to exact solvability. In the special case of $\{0,1\}$-valued constraint languages,
which correspond to Max-CSPs, constraint languages on two-element
domains~\cite{Creignou95:maximum}, three-element
domains~\cite{JonssonKK06:3valued}, four-element
domains~\cite{Jonsson11:cp}, and conservative (containing all $\{0,1\}$-valued
unary functions) constraint languages~\cite{DeinekoJKK08} have been classified with respect
to exact solvability.
Generalising the algebraic approach to CSPs~\cite{Bulatov05:classifying},
algebraic properties called multimorphisms~\cite{Cohen06:complexitysoft},
fractional polymorphisms~\cite{Cohen06:expressive}, and weighted
polymorphisms~\cite{cccjz13:sicomp} have been invented for
the study of the computational complexity of classes of VCSPs.

\subsection{Contribution}
We study the computational complexity of finite-valued constraint languages on
arbitrary finite domains. We characterise all tractable finite-valued constraint
languages as those admitting a binary symmetric fractional polymorphism.
Tractability follows from the results
in~\cite{tz12:focs,Kolmogorov13:icalp} (see also~\cite{ktz15:sicomp}, which
is an extended version of~\cite{tz12:focs} and~\cite{Kolmogorov13:icalp}) that show that all
instances over such constraint languages are solvable by the basic linear
programming relaxation (BLP). In the other direction, we show that instances
over constraint languages not admitting such a fractional polymorphism are
NP-hard by a reduction from Max-Cut~\cite{Garey79:intractability}. 

\begin{theorem}\label{thm:main}
  Let $D$ be an arbitrary finite set and let $\Gamma$ be a finite-valued
  constraint language defined on $D$.
  VCSP$(\Gamma)$ is tractable if, and only if, the BLP solves VCSP$(\Gamma)$.
  Otherwise, VCSP$(\Gamma)$ is NP-hard.
\end{theorem} 

An explicit hardness condition is given in Theorem~\ref{thm:dichotomy}.

Our results generalise \emph{all previous partial classifications} of
finite-valued constraint languages: the classifications of $\{0,1\}$-valued
constraint languages on
two-element, three-element, and four-element domains obtained
in~\cite{Creignou95:maximum,Creignouetal:siam01}, \cite{JonssonKK06:3valued},
and~\cite{Jonsson11:cp}, respectively; the classification of $\{0,1\}$-valued
constraint languages containing all unary functions obtained in~\cite{DeinekoJKK08}; the
classifications of finite-valued constraint languages on two-element and three-element
domains obtained in~\cite{Cohen06:complexitysoft} and~\cite{hkp14:sicomp},
respectively; the classification of finite-valued constraint languages containing all
$\{0,1\}$-valued unary functions obtained in~\cite{kz13:jacm}; and the
classification of Min-0-Ext problems obtained in~\cite{hirai15:mp}.

Our results demonstrate that (i) a binary symmetric fractional polymorphism
is sufficient for characterising tractability, and (ii) only cores and constants
are required for the hardness condition (details are explained in
Section~\ref{sec:prelim}). 
This is in contrast with ordinary CSPs (that
is, the decision problems), where the hardness condition also requires an
equivalence relation and the conjectured tractable cases are characterised by
polymorphisms of arity higher than two~\cite{Bulatov05:classifying}.

Another problem tackled here is referred to,
in~\cite{Creignouetal:siam01}, as the \emph{meta problem}: given
a finite-valued constraint language $\Gamma$, decide whether it gives rise to a
tractable class VCSP$(\Gamma)$. We show that the meta problem is solvable in
polynomial time when the constraint language $\Gamma$ is assumed to be a core.
However, we also show that deciding whether $\Gamma$ is a core is co-NP-complete
and that deciding whether a given $\Gamma'$ is a core of $\Gamma$ is
DP-complete. In particular, all considered meta problems are decidable.

A finite-valued constraint language $\Gamma$ is called \emph{tractable} if every
finite subset $\Gamma'\subseteq\Gamma$ gives rise to a tractable class
VCSP$(\Gamma')$.
However, in principle, the algorithms solving VCSP$(\Gamma')$
for different finite subsets of $\Gamma$ could be quite different. If there
exists a uniform polynomial-time algorithm for VCSP($\Gamma$) then we say that
$\Gamma$ is \emph{globally tractable}. In the case of ordinary CSPs
(that is, decision problems), in all known cases every tractable constraint
language is also globally tractable.
%
%
Our results show that this holds in general for finite-valued constraint languages:
all tractable infinite constraint languages are globally tractable, using the BLP
relaxation, and all other constraint languages are NP-hard.
We therefore derive a dichotomy result also for \emph{infinite} finite-valued
constraint languages.

The proof of our main result is a combination of various techniques.
We elaborate on a slightly different, but equivalent, notion of core for
finite-valued constraint languages from that used in~\cite{hkp14:sicomp}.
We introduce the idea of studying expressible unary functions by encoding them
in hyperplane arrangements.
We also use the idea introduced in~\cite{Kolmogorov13:icalp} of working
with \emph{generalised fractional polymorphisms} but derive the necessary
technical machinery using a Markov chain argument.
This also provides
natural way to derive the main result from~\cite{Kolmogorov13:icalp} which says that
having a binary symmetric fractional polymorphism implies having symmetric
fractional polymorphisms of all arities.

Since the announcement of our results in the conference version of this
article~\cite{tz13:stoc}, the techniques presented here have proved essential in
recent complexity classifications of Min-Sol problems and Min-Cost-Hom problems,
which are special cases of VCSPs~\cite{Uppman13:icalp,Uppman14:stacs}.

\subsection{Related work}
Apart from language-based restrictions on (V)CSPs, also structure-based
restrictions~\cite{Grohe07:compl,Marx13:jacm,Gottlob09:icalp,Farnqvist12:cpaior}
and hybrid restrictions~\cite{cz11:ai,cz12:jair} have been studied.
Not only exact solvability, but also approximability of Max-CSPs and VCSPs has
attracted a lot of
attention~\cite{Creignouetal:siam01,Khanna01:approximability,Hastad01:jacm,Hastad08:every,Guruswami08:approx-random,Jonsson09:tcs}.
Moreover, the robust approximability of Max-CSPs has also been
studied~\cite{Kun12:itcs,Barto12:stoc,Dalmau13:robust}. 
Under the assumption of the \emph{unique games conjecture}~\cite{khot10:coco},
Raghavendra has shown that the basic semidefinite programming (SDP) relaxation
solves all tractable finite-valued CSPs (without a characterisation of the
tractable cases)~\cite{Raghavendra08:stoc}. Moreover, Chapters~6 and~7
of~\cite{Raghavendra09:phd} imply that if a finite-valued constraint language
$\Gamma$ admits a cyclic fractional polymorphism of some arity $k\geq 2$ then
the basic SDP relaxation solves any VCSP instance over $\Gamma$. Our results
show, assuming P $\neq$ NP, that for exact solvability the BLP relaxation
suffices.

\section{Preliminaries}\label{sec:prelim}

We use the following notation: any name with a bar denotes a tuple. We denote by
$x_i$ the $i$th component of a tuple $\tup{x}$. Superscripts are used for
collections of tuples; e.g., we write $x^j_i$ for the $i$th component of the $j$th tuple $\tup{x}^j$.

\subsection{Valued CSPs}
Let $D$ be a finite set called the \emph{domain}.
We denote by $\Qp$, $\Qnn$, and $\Q$, respectively, the set of positive rational
numbers, nonnegative rational numbers, and rational numbers. 
A \emph{(cost) function} is any function
$f:D^m\to\range$, where $m=ar(f)$ is the \emph{arity} of $f$.
A \emph{valued constraint language} $\Gamma$ is a set of cost functions.
Unless specifically said otherwise, we assume that all constraint languages under
consideration are \emph{finite}.
Valued
constraint languages consisting of $\Q$-valued cost functions that do not take on
infinite costs are called \emph{finite-valued} constraint languages in the literature and this
is the term we used in the abstract and introduction. Since we exclusively study
finite-valued constraint languages, for simplicity we omit the words ``valued'' and
``finite-valued'' and in the rest of the paper we say simply ``constraint
language''.

\begin{definition}\label{def:vcsp}
An instance $I$ of the \emph{valued constraint satisfaction problem}, or VCSP
for short,
is
given by the set $V=\{x_1,\ldots,x_n\}$ of variables and the objective function
$f_I(x_1,\ldots,x_n)=\sum_{i=1}^qw_i\cdot f_i(\tup{x}^i)$ where, for every
$1\leq i\leq q$,
$f_i:D^{ar(f_i)}\to\range$, $\tup{x}^i\in V^{ar(f_i)}$, and
$w_i\in\Qnn$ is a weight.
The functions $f_i$ are extensionally represented, i.e.,
given by a table of costs for all possible $|D|^{ar(f_i)}$ assignments.
A solution to $I$ is a function $h : V \to D$, its measure given by
$\sum_{i=1}^q w_i\cdot f_i(h(\tup{x}^i))$,
where $h$ is applied componentwise.
The goal is to find a solution of minimum measure. 
\end{definition}
We denote by VCSP$(\Gamma)$ the class of all instances in which all functions
are from $\Gamma$.
The minimum measure of a solution to an instance $I\in$ VCSP$(\Gamma)$ is denoted
by $\opt_\Gamma(I)$.
A constraint language $\Gamma$ is called \emph{tractable} if, for any finite
$\Gamma'\subseteq \Gamma$, VCSP$(\Gamma')$ is tractable, that is, a
solution of measure $\opt_\Gamma(I)$ can be found for any instance $I\in$ VCSP$(\Gamma')$
in polynomial time; $\Gamma$ is called \emph{NP-hard} if VCSP$(\Gamma')$ is
NP-hard for some finite $\Gamma'\subseteq\Gamma$. Moreover, $\Gamma$ is called
\emph{globally tractable} if there is a uniform algorithm for VCSP$(\Gamma)$.

\subsection{Expressive power}\label{sec:express}

\begin{definition}
For a constraint language $\Gamma$, we let $\express{\Gamma}$ be the set of all
functions $f(x_1,\ldots,x_m)$ such that for some instance $I\in$ VCSP$(\Gamma)$
with objective function 
  $f_I(x_1, \ldots, x_m, x_{m+1}, \ldots, x_n)$, we have
\[ f(x_1,\ldots,x_m)\ =\
\min_{x_{m+1},\ldots,x_n}f_I(x_1,\ldots,x_m,x_{m+1},\ldots,x_n)\,. \]
We then say that $\Gamma$ \emph{expresses} $f$ and call $\express{\Gamma}$ the
\emph{expressive power} of $\Gamma$. 
\end{definition}
In other words, $\express{\Gamma}$ is the closure of
$\Gamma$ under addition, multiplication by nonnegative constants,
and minimisation over extra variables.
For two functions $f$
and $f'$, we write $f\equiv f'$ if $f=a\cdot f'+b$ for some $a\in\Q_{>0}$
and $b\in\Q$, i.e., 
if $f$ can be obtained from
$f'$ by \emph{scaling} and \emph{translation}.
For a constraint language $\Gamma$, let
$\Gamma_\equiv=\{f\:\mid\:f\equiv f'\mbox{ for some }f'\in \Gamma\}$.
It has been shown that with respect to exact solvability, we only need to
consider constraint languages closed under expressibility,
scaling, and translation:

\begin{theorem}[\cite{Cohen06:complexitysoft}]\label{thm:expr}
Let $\Gamma$ be a constraint language and $\Gamma'$ a finite set
such that
$\Gamma'\subseteq\express{\Gamma}_\equiv$.
Then VCSP$(\Gamma')$ polynomial-time reduces to VCSP$(\Gamma)$.
\end{theorem}
\noindent
We define the following condition:
\begin{quote}{MC}
{\em There exist distinct $a,b \in D$ such that
$\express{\Gamma}$ contains 
a binary function $f$ with $\argmin f= \{(a,b),(b,a)\}$.}
\end{quote}
A slightly different condition\footnote{Condition (MC$'$) was called (MC)
in~\cite{hkp14:sicomp}.} was formulated in~\cite{hkp14:sicomp}:
 \begin{quote}{MC$'$}
 {\em There exist distinct $a,b \in D$ such that $\express{\Gamma}$
contains a unary function $u$ with $\argmin u = \{a,b\}$
and a binary function $f$ with
$f(a,b) = f(b,a) < f(a,a) = f(b,b)$.}
 \end{quote}
Observe that (MC$'$) implies (MC).
In fact, we will now prove that the two conditions are equivalent.
\begin{lemma}
\label{lem:mctomcp}
  For any constraint language $\Gamma$, (MC) holds if, and only, if (MC\,$'$) holds.
\end{lemma}

\begin{proof}
  We need to prove that (MC) implies (MC$'$).
  Let $\Gamma$ be a constraint language with a function $f \in \express{\Gamma}$ such that $\argmin f = \{(a,b),(b,a)\}$.
  Note that $u(x) = \min_y f(x,y)$ is a unary function with $\argmin u = \{a,b\}$.
  If $f(a,a) = f(b,b)$, then $u$ and $f$ satisfy (MC$'$).
  Otherwise, 
  assume without loss of generality that $f(a,b) = f(b,a) = 0$,
  $f(x,y) \geq 1$ for $\{x,y\} \neq \{a,b\}$,
  and that
  $f(a,a) < f(b,b)$.
  Let $K = \max_x f(a,a)-f(x,x)$, and define
  $u'(x) = \min_y K \cdot u(y) + f(y,y) + f(x,y)$.
  Note that $u(x) = 0$ for $x = a, b$ and $u(x) \geq 1$ otherwise.
  Also note that $\min_{y} f(y,y) = f(a,a)-K$.
  The three arguments in the following $\min$-expressions correspond to the cases
  $y \not\in\{a,b\}$, $y = a$, and $y = b$, respectively.
  \begin{align*}
    u'(x) & \geq \min \{K+(f(a,a)-K)+1, 0+f(a,a)+1, 0+f(b,b)+1 \} > f(a,a) \qquad (x \neq a,b) \\
    u'(a) & \geq \min \{K+(f(a,a)-K)+1, 0+f(a,a)+f(a,a), 0+f(b,b)+0\} > f(a,a) \\
    u'(b) & \leq K \cdot f(a,b) + f(a,a) + f(b,a) = f(a,a)
  \end{align*}
  Thus $\argmin u' = \{b\}$.

  Now, let $\delta = f(b,b)-f(a,a) > 0$ and define
  $$f'(x,y) = f(x,y) + \frac{\delta}{2}\frac{u'(x)+u'(y)}{u'(a)-u'(b)}.$$
  We now verify that $f'(a,b) = f'(b,a) < f'(a,a) = f'(b,b)$:
   \begin{align*}
     f'(a,a)-f'(a,b) & =  f(a,a) + \frac{\delta}{2}\frac{u'(a)+u'(a)}{u'(a)-u'(b)}
     - f(a,b) - \frac{\delta}{2}\frac{u'(a)+u'(b)}{u'(a)-u'(b)} \\
     & = f(a,a) + (f(b,b)-f(a,a))\frac{1}{2} \frac{u'(a)-u'(b)}{u'(a)-u'(b)} \\
     & = \frac{1}{2} (f(a,a)+f(b,b)) > 0, \\
     f'(a,a)-f'(b,b) & = f(a,a) + \frac{\delta}{2}\frac{u'(a)+u'(a)}{u'(a)-u'(b)}
     - f(b,b) - \frac{\delta}{2}\frac{u'(b)+u'(b)}{u'(a)-u'(b)} \\
     & = f(a,a)-f(b,b) + (f(b,b)-f(a,a))\frac{u'(a)-u'(b)}{u'(a)-u'(b)} = 0, \\
     f'(a,b)-f'(b,a) & = f(a,b) + \frac{\delta}{2}\frac{u'(a)+u'(b)}{u'(a)-u'(b)}
     - f(b,a) - \frac{\delta}{2}\frac{u'(b)+u'(a)}{u'(a)-u'(b)} = 0.
   \end{align*}
   It follows that $u$ and $f'$ satisfy (MC$'$).
\end{proof}

It is known that condition (MC$'$) and thus, by Lemma~\ref{lem:mctomcp},
condition (MC) implies intractability (via a reduction from
Max-Cut~\cite{Garey79:intractability}):
\begin{lemma}[\cite{Cohen06:complexitysoft}]\label{lem:mc}
If a constraint language $\Gamma$ satisfies condition $(MC)$ then $\Gamma$ is
NP-hard.
\end{lemma}

\subsection{Fractional polymorphisms}
For a cost function
$f$ and $\tup{a}^1,\dots,\tup{a}^m\in D^{ar(f)}$, let
$f^m(\tup{a}^1,\ldots,\tup{a}^m):=\frac{1}{m}(f(\tup{a}^1)+\dots+f(\tup{a}^m))$.
An $m$-ary
\emph{operation} on $D$ is a function $g : D^m \rightarrow D$. Let $\opm{D}{m}$
denote the set of all $m$-ary operations on $D$. 
An $m$-ary \emph{fractional
operation} is a function $\omega : \opm{D}{m} \rightarrow \Qnn$
such that $\|\omega\|_1 = 1$,
where $\|\omega\|_1 := \sum_{g} \omega(g)$.\footnote{In~\cite{tz13:stoc}, fractional operations were defined without the requirement $\|\omega\|_1=1$ which was instead added to the definition of fractional polymorphisms. The present definition better matches the semantics of the qualifier ``fractional''.}
The set $\{ g \mid \omega(g) > 0 \}$ of operations is called the \emph{support}
of $\omega$ and is denoted by $\supp(\omega)$.
For an operation $g$, we denote by $\chi_g$ the fractional operation that
takes the value 1 on the operation $g$ and 0 on all other operations.

A fractional operation $\omega$
is called an $m$-ary \emph{fractional polymorphism}~\cite{Cohen06:expressive} of
$f$ if, for all
${\tup{a}^1}, \dots, \tup{a}^m \in D^{ar(f)}$, it holds that
\begin{equation}\label{ineq:fpol}
\sum_{g \in \opm{D}{m}} \omega(g) f(g(\tup{a}^1,\dots,\tup{a}^m)) \leq
f^m(\tup{a}^1,\ldots,\tup{a}^m),
\end{equation}
where the operations $g$ are applied componentwise.
If $\omega$ is a fractional polymorphism of $f$ then we say that $\omega$
\emph{improves} $f$ and that $f$ \emph{admits} the fractional polymorphism
$\omega$. 

If $\omega$ is a fractional polymorphism of every cost function in a constraint language $\Gamma$,
then $\omega$ is called a fractional polymorphism of $\Gamma$, and we say that
$\Gamma$ admits the fractional polymorphism $\omega$.

It is known and easy to show that
expressibility preserves fractional polymorphisms: 
if $\omega$ is a fractional polymorphism of $\Gamma$ then $\omega$ is also a
fractional polymorphism of $\express{\Gamma}$~\cite{Cohen06:expressive}.

An operation $g$ is \emph{idempotent} if $g(x,\dots,x)=x$. Let $S_m$ be the
symmetric group on $\{1,\dots,m\}$. An $m$-ary operation $g$ is \emph{symmetric}
if, for every permutation $\pi \in S_m$, we have $ g(x_1,\dots,x_m) =
g(x_{\pi(1)},\dots,x_{\pi(m)}).$ An $m$-ary operation $g$ is \emph{cyclic} if
$g(x_1,x_2,\ldots,x_m)=g(x_2,\ldots,x_m,x_1)$ for all $x_1,\ldots,x_m\in D$.
Note that in the case of $m=2$, an operation is symmetric if, and only if,
it is cyclic. 
A fractional operation is called idempotent, symmetric, or cyclic if all
operations in its support are idempotent, symmetric, or cyclic, respectively.

A \emph{mapping} of arity $m\to k$ on $D$ is a function $\mathbf{g} : D^m \to D^k$.
Let $\opm{D}{m\rightarrow k}$ denote the set of all mappings of arity $m\to k$ on $D$.
A \emph{fractional mapping} (of arity $m\to k$) is a function $\rho :
\opm{D}{m\rightarrow k} \rightarrow \Qnn$ such that
$\|\rho\|_1 = 1$, where $\|\rho\|_1 := \sum_{\mathbf{g}} \rho(\mathbf{g})$. 
A fractional mapping $\rho$ is called a \emph{generalised fractional
polymorphism}
(of arity $m\to k$) of $f$ if, for all $\tup{a}^1,
\dots, \tup{a}^m \in D^{ar(f)}$, it holds that
\begin{equation}
\sum_{\mathbf{g}\in \opm{D}{m\rightarrow k}} \rho(\mathbf{g})
f^k(\mathbf{g}(\tup{a}^1,\dots,\tup{a}^m)) \leq f^m(\tup{a}^1,\ldots,\tup{a}^m).
\end{equation}

As for ordinary fractional polymorphisms, we say that $\rho$ is a generalised
fractional polymorphism of a constraint language $\Gamma$ if $\rho$ is a
generalised fractional polymorphism of every cost function from $\Gamma$ and say
that $\Gamma$ admits $\rho$.

The definitions of the fractional mapping $\chi_{\mathbf{g}}$, given a mapping
$\mathbf{g}$, and of the support $\supp(\rho)$ of a fractional mapping $\rho$
are analogous to those for fractional operations.

A mapping $\mathbf{g}$ of arity $m \to k$ is \emph{symmetric} if,
for every permutation $\pi \in S_m$, we have
$\mathbf{g}(x_1,\dots,x_m) = \mathbf{g}(x_{\pi(1)},\dots,x_{\pi(m)})$,
and a fractional mapping is called symmetric if all
mappings in its support are symmetric.

Note that a fractional polymorphism of arity $m$ is the same as a generalised
fractional polymorphism of arity $m\rightarrow 1$.
In fact a fractional mapping of arity $m\rightarrow k$ is just a
tuple of $k$ fractional operations of arity $m\rightarrow 1$; however, this
viewpoint, introduced in~\cite{Kolmogorov13:icalp}, turns out to be very useful.
For brevity, we will often omit the word ``generalised'' when no ambiguity can arise.

\subsection{Cores}\label{sec:cores}

Let $S \subseteq D$.
The \emph{sub-language $\Gamma[S]$ of $\Gamma$ induced by $S$} is the constraint
language defined on domain $S$ and containing the restriction of every function
$f\in\Gamma$ onto $S$.

\begin{definition}
\label{def:core}
  A constraint language $\Gamma$ is a \emph{core} if for every unary fractional
  polymorphism $\omega$ of $\Gamma$, $\supp(\omega)$ contains only injective
  operations.
  A constraint language $\Gamma'$ is a \emph{core of $\Gamma$} if $\Gamma'$ is a
  core and $\Gamma' = \Gamma[g(D)]$ for some $g \in \supp(\omega)$ with $\omega$
  a unary fractional polymorphism of $\Gamma$.
\end{definition}

The following lemma implies that we may always assume that $\Gamma$ is a core constraint language.
It is an immediate consequence of Lemma~\ref{lem:operationsinomega} below.

\begin{lemma}\label{lem:core}
  If $\Gamma'$ is a core of $\Gamma$ then $\opt_\Gamma(I) = \opt_{\Gamma'}(I')$
  for all instances $I\in$ VCSP$(\Gamma)$, where $I'$ is obtained from $I$ by
  substituting each function in $\Gamma$ for its restriction in $\Gamma'$.
\end{lemma}

We will need the following variation of Motzkin's transposition theorem.
\begin{lemma}
\label{lem:motzkin}
  For any
  $A \in \Q^{m \times n}$,
  $B \in \Q^{p \times n}$,
  exactly one of the following holds:
  \begin{itemize}
  \item
    $Ay > 0$, $By \geq 0$, for some $y \in \Q_{\geq 0}^n$; or
  \item
    $A^{\top}z_1 + B^{\top}z_2 \leq 0$, for some $0 \neq z_1 \in \Q_{\geq 0}^m, z_2 \in \Q_{\geq 0}^p$.
  \end{itemize}
\end{lemma}
\begin{proof}
The following variation of Motzkin's transposition theorem is from~\cite[Corollary~7.1k]{Schrijver86:ILP} with $b=c=0$ and the matrices multiplied by $-1$:
For any
  $A' \in \Q^{m' \times n'}$,
  $B' \in \Q^{p' \times n'}$,
  exactly one of the following holds:
  \begin{enumerate}
  \item \label{M1}
    $A'y' > 0$, $B'y' \geq 0$, for some $y' \in \Q^{n'}$; or
  \item \label{M2}
    $A'^{\top}z'_1 + B'^{\top}z'_2 = 0$, for some $0 \neq z'_1 \in \Q_{\geq 0}^{m'}, z'_2 \in \Q_{\geq 0}^{p'}$.
  \end{enumerate}

Given $A$ and $B$ as in the statement of the lemma, set 
$n' = n$, $m' = m$, $p' = p+n$,
$A'=A$ and
$B'=\left(\begin{array}{c} B \\ I_{n\times n}\end{array}\right)$, where
$I_{n\times n}\in\Q^{n\times n}$ is the identity matrix. 

Firstly, observe that~(\ref{M1}), i.e., the existence of some $y'\in\Q^{n'}$ satisfying
$A'y>0$ and $B'y\geq 0$, is equivalent to the first case of the lemma, i.e., the
existence of some $y\in\Q^n_{\geq 0}$ satisfying $Ay>0$ and $By\geq 0$. 
Secondly, observe that~(\ref{M2}), i.e., the existence of some
$0 \neq z'_1 \in \Q_{\geq 0}^{m'}, z'_2 \in \Q_{\geq 0}^{p'}$ satisfying $A'^{\top}z'_1 +
B'^{\top}z'_2 = 0$, is equivalent to the second case of the lemma, i.e., the
existence of some
$0 \neq z_1 \in \Q_{\geq 0}^m, z_2 \in \Q_{\geq 0}^p$ satisfying
$A^{\top}z_1 + B^{\top}z_2 \leq 0$. To see this, note that the last $n$ coordinates
of $z'_2$ can be independently chosen, and therefore set to satisfy $A'^{\top}z'_1 +
B'^{\top}z'_2 = 0$ as long as $A^{\top}z_1 + B^{\top}z_2 \leq 0$.
This shows that~(\ref{M2}) is implied by the second case of
the lemma, and the other direction holds trivially.
\end{proof}
\begin{lemma}
\label{lem:operationsinomega}
  For a constraint language $\Gamma$,
  and a unary operation $g \in \opm{D}{1}$,
  the following are equivalent:
  \begin{enumerate}
  \item\label{item:op1}
    $\Gamma$ admits a unary fractional polymorphism $\omega$ with $g \in \supp(\omega)$.
  \item\label{item:op2}
    For all instances $I$ of VCSP$(\Gamma)$ and all optimal solutions $s$ to $I$, $g \circ s$ is also an optimal solution to $I$.
  \end{enumerate}
\end{lemma}
\begin{proof}
The first condition of the lemma holds if and only if the following
system of linear inequalities is satisfiable:
\begin{equation}
\begin{aligned}
  \sum_{h \in \opm{D}{1}} \omega(h) f(h(\tup{x})) & \leq \|\omega\|_1
  f(\tup{x}) \quad \text{ $\forall f\in\Gamma, \tup{x}\in D^{ar(f)}$} \\ 
  \omega(g) & > 0 \\ 
  \omega(h) & \geq 0 \quad \text{ $\forall h\in\opm{D}{1}$}. 
\label{eq:op}
\end{aligned}
\end{equation}
  According to Lemma~\ref{lem:motzkin}, this is true if, and only if, the
  following system is unsatisfiable:
  \begin{equation}
  \begin{aligned}
    \sum_{f \in \Gamma,\tup{x} \in D^{ar(f)}} z_2(f,\tup{x}) (f(\tup{x})-f(h(\tup{x}))) & \leq 0, \quad \text{ $\forall h \in \opm{D}{1}$}, \\
    z_1 + \sum_{f \in \Gamma,\tup{x} \in D^{ar(f)}} z_2(f,\tup{x}) (f(\tup{x})-f(g(\tup{x}))) & \leq 0, \\
    z_1 & > 0, \\
    z_2(f,\tup{x}) & \geq 0, \quad \text{ $\forall f \in \Gamma, \tup{x} \in D^{ar(f)}$}. \label{eq:op2}
  \end{aligned}
  \end{equation}
  Let $V_D = \{ v_a \mid a \in D \}$ and define $\iota : V_D \to D$ by $\iota(v_a) = a$.
  Then, (\ref{eq:op2}) is unsatisfiable if, and only if, there is no instance
  $J$ of VCSP$(\Gamma)$,
  with variables $V(J) = V_D$ and objective function 
  $f_J = \sum_{f,\tup{x}} z_2(f,\tup{x}) f(\iota^{-1}(\tup{x}))$
  such that $g \circ \iota$ is a non-optimal solution.

  It is clear that the second condition of the lemma implies that (\ref{eq:op2}) is unsatisfiable.
  It remains to show the reverse implication.
  Let $I$ be any instance of VCSP$(\Gamma)$ and $s : V(I) \to D$ any optimal solution to $I$.
  Construct an instance $J$ of VCSP$(\Gamma)$ with variables $V(J) = V_D$ 
  by replacing each term $w_i \cdot f_i(\tup{x}^i)$ in $f_I$ by the term $w_i \cdot f_i(\iota^{-1} \circ s(\tup{x}^i))$ in $f_J$.
  Since (\ref{eq:op2}) is unsatisfiable, it follows that $g \circ \iota$ is an optimal solution to $J$, and
  hence that $g \circ s$ is an optimal solution to $I$.
  As $I$ and $s$ were chosen arbitrarily,
  this establishes the lemma.
\end{proof}

In~\cite{hkp14:sicomp}, a constraint language $\Gamma$ is defined to be a core if,
for each $a \in D$, there is an instance $I_a$ of VCSP$(\Gamma)$ such that
$a$ appears in every optimal solution to $I_a$.
We now show that this condition is equivalent to Definition~\ref{def:core}.

\begin{lemma}
\label{lem:ourcore-theircore}
  For a constraint language $\Gamma$,
  the following are equivalent:
  \begin{enumerate}
  \item\label{item:core1}
    All unary fractional polymorphisms of $\Gamma$ are injective.
  \item\label{item:core3}
    For each $a \in D$, there is an instance $I_a$ of VCSP$(\Gamma)$ such that
    $a$ appears in every optimal solution to $I_a$.
  \end{enumerate}
\end{lemma}
\begin{proof}
First we show the implication $(2)\Rightarrow (1)$.
Assume that $(1)$ does not hold and
let $\omega$ be a unary fractional
polymorphism of $\Gamma$ with a non-injective $g\in\supp(\omega)$; that is,
there is an $a\in D$ such that $a\not\in g(D)$.
Then, Lemma~\ref{lem:operationsinomega} implies that every instance of VCSP$(\Gamma)$
has a solution where $a$ does not appear, so $(2)$ does not hold. 

We now show $(1)\Rightarrow (2)$.
  By Lemma~\ref{lem:operationsinomega}, condition (\ref{item:core1}) holds if, and only if,
  for every non-injective unary operation $g \in \opm{D}{1}$,
  there exists an instance $I_g$ of VCSP$(\Gamma)$ and an optimal solution $s_g$ to $I_g$
  such that $g \circ s_g$ is not an optimal solution to $I_g$.
  Let $f_{I_g} = \sum_i w_i \cdot f_i(\tup{x}^i)$ be the objective function of $I_g$, and,
  as in the proof of Lemma~\ref{lem:operationsinomega}, construct an instance $J_g$ with
  variables $V_D = \{ v_a \mid a \in D\}$ and objective function $f_{J_g} = \sum_i w_i \cdot f_i(\iota^{-1}(\tup{x}^i))$, where $\iota : V_D \to D$ given by $\iota(v_a) = a$.
  Then, $\iota$ is an optimal solution to $J_g$, but $g \circ \iota$ is not.
  Let $I$ be the instance with variables $V_D$ and $f_I = \sum_g f_{J_g}$,
  where the sum is over all non-injective unary operations.
  Let $s$ be an optimal solution to $I$.
  Note that $s$ must also be an optimal solution to each instance $J_g$.
  Since $s \circ \iota^{-1}$ is a unary operation on $D$, it follows that $s$ must be injective,
  hence for every $a \in D$, there is a $v \in V_D$ such that $s(v) = a$.
  We can therefore let $I_a := I$ for each $a \in D$.
\end{proof}

For a constraint language $\Gamma$, let $\Gamma_c$ denote the set of all functions obtained from functions in $\Gamma$ by fixing a (possibly empty)
subset of the variables to domain values.
We will use the following result, which says that 
we can restrict our attention to core constraint languages whose expressive powers
contain certain unary functions.
\begin{proposition}[\cite{hkp14:sicomp}]\label{prop:core}
  Let $\Gamma$ be a core constraint language defined on a finite domain $D$.
\begin{enumerate}

\item For each $a \in D$, $\express{\Gamma_c}$ contains a unary function $u_a$ such that $\argmin u_a = a$.

\ignore{
\item $\Gamma$ is tractable if, and only if, $\Gamma_c\cup\{u_a\:|\:a\in D\}$ is
tractable.

\item $\Gamma$ is NP-hard if, and only if, $\Gamma_c\cup\{u_a\:|\:a\in D\}$ is
NP-hard. \label{core:hardness}
}

\item $\Gamma$ is NP-hard if, and only if, $\Gamma_c$ is NP-hard. \label{core:hardness}
\end{enumerate}
\end{proposition}

It follows readily from Proposition~\ref{prop:core} that every (generalised)
fractional polymorphism of $\Gamma_c$ for a core constraint language $\Gamma$ is
idempotent.

\section{Complexity classification}\label{sec:results}

The computational complexity of constraint languages has attracted a lot
of attention in the literature. The partial classifications obtained before the results
of this paper can be summarised as follows:
\begin{itemize}

\item $\{0,1\}$-valued constraint languages on $|D|=2$~\cite{Creignou95:maximum,Creignouetal:siam01}.
\item $\{0,1\}$-valued constraint languages on $|D|=3$~\cite{JonssonKK06:3valued}.
\item $\{0,1\}$-valued constraint languages on $|D|=4$~\cite{Jonsson11:cp}.
\item $\{0,1\}$-valued constraint languages containing all $\{0,1\}$-valued unary functions~\cite{DeinekoJKK08}.
\item constraint languages on $|D|=2$~\cite{Cohen06:complexitysoft}.
\item constraint languages on $|D|=3$~\cite{hkp14:sicomp}.
\item constraint languages containing $\{0,1\}$-valued unary functions~\cite{kz13:jacm}.
\item constraint languages containing unary functions and certain special binary
functions~\cite{hirai15:mp}.
\end{itemize}

In all of these classifications, the hardness reductions essentially came from
the condition (MC) and tractable cases were characterised by certain specific
binary symmetric fractional polymorphisms including the concepts of
submodularity~\cite{JonssonKK06:3valued,DeinekoJKK08,Cohen06:complexitysoft}, skew
bisubmodularity~\cite{hkp14:sicomp}, 1-defect~\cite{Jonsson11:cp}, and
others~\cite{hirai15:mp}.

\subsection{The basic linear programming relaxation}

Every VCSP instance has a natural linear programming relaxation, proposed
independently by a number of
authors~\cite{Schlesinger76,Koster98,Chekuri04:sidma,Wainwright05,Carleton05,Cooper08:minimizing,Cooper10:osac,Kun12:itcs}.
This relaxation is referred to as the \emph{basic LP relaxation} (BLP)
as it is the first level in the Sherali-Adams hierarchy \cite{Sherali1990}.
It can be defined as follows.

Let $\Gamma$ be a 
constraint language defined on $D$
and let $I$ be a VCSP$(\Gamma)$ instance given by
the set $V=\{x_1,\ldots,x_n\}$ of variables and the objective function
$f_I(x_1,\ldots,x_n)=\sum_{i=1}^qw_i\cdot f_i(\tup{x}^i)$ where, for every
$1\leq i\leq q$,
$f_i:D^{ar(f_i)}\to\range$, $\tup{x}^i\in V^{ar(f_i)}$, and
$w_i\in\Qnn$ is a weight.
For a tuple $\tup{x}$, let $\{\tup{x}\}$ denote the set of elements in
$\tup{x}$.
The BLP has variables $\lambda_{i,\sigma_i}$,
for $1\leq i\leq q$ and $\sigma_i : \{{\tup{x}^i}\} \rightarrow D$; 
and variables $\mu_{x,a}$, for $x \in V$ and $a \in D$.
\begin{equation*}
  \begin{array}{lll}
    \min
& \multicolumn{2}{l}{\displaystyle \sum_{i=1}^q\ w_i \sum_{\sigma_i :
\{{\tup{x}^i}\} \rightarrow D} f_i(\sigma_i({\tup{x}^i}))\cdot
\lambda_{i,\sigma_i}} \\
    \text{s.t.}
& \displaystyle\sum_{\substack{\sigma_i : \{\tup{x}^i\}\rightarrow
D\\\sigma_i(x) = a}}
\lambda_{i,\sigma_i} = \mu_{x,a} 
& \text{$\forall 1\leq i\leq q, \forall x\in \{\tup{x}^i \},
\forall a\in D$} \\
& \hspace*{0.8em} \displaystyle\sum_{a \in D} \mu_{x,a} = 1 
& \text{$\forall x \in V$} \\
\smallskip & 
\quad 0 \leq \lambda, \mu \leq 1
  \end{array}
\end{equation*}

Since $\Gamma$ is fixed, this relaxation has polynomial size in $I$.
Requiring
$\lambda_{i,\sigma_i}$ and $\mu_{x,a}$ to be in $\{0,1\}$ provides an integer
programming formulation of $I$ with the meaning $\mu_{x,a}=1$ if, and only if,
variable $x$ is assigned value $a$.

For any VCSP instance $I$, the BLP gives a lower bound on the measure of an
optimal solution to $I$. Denote this lower bound by BLP$(I)$.
We will say that the BLP \emph{solves} VCSP$(\Gamma)$ if BLP$(I) = \opt_\Gamma(I)$
for every $I \in $ VCSP$(\Gamma)$.
It can be shown that when the BLP solves VCSP$(\Gamma)$, then a solution attaining
the optimum can also be obtained in polynomial time~\cite{ktz15:sicomp}.

A result of the authors characterised the constraint languages for
which the BLP relaxation solves VCSP$(\Gamma)$
in terms of symmetric fractional polymorphisms~\cite{tz12:focs}.
An equivalent simplified condition was subsequently given in~\cite{Kolmogorov13:icalp}, see
also~\cite{ktz15:sicomp}.

\begin{theorem}[\cite{tz12:focs,Kolmogorov13:icalp}]\label{thm:blp}
Let $\Gamma$ be a constraint language. Then BLP solves VCSP$(\Gamma)$ if, and only
if, $\Gamma$ admits a binary symmetric fractional polymorphism.
\end{theorem}

\subsection{Main classification}
\label{subsec:main}

The main technical contribution of this paper is the following result.

\begin{restatable}{theorem}{mainimpl}
\label{thm:mainimpl}
Let $D$ be an arbitrary finite set and let $\Gamma$ be a 
constraint language defined on $D$.
If $\Gamma$ is a core such that $\Gamma_c$ does not satisfy (MC), then $\Gamma$
admits a binary idempotent and symmetric fractional polymorphism.
\end{restatable}

We will also need the following lemma which is proved in Section~\ref{sec:newlemmaproof}.

\begin{lemma}
\label{lem:symcore}
Let $\Gamma$ be a constraint language defined on $D$ and let $\Gamma'$ be a core of $\Gamma$.
If $\Gamma'$ admits a binary symmetric
fractional polymorphism, then so does $\Gamma$.
\end{lemma}

Theorem~\ref{thm:mainimpl} implies our main result,
Theorem~\ref{thm:dichotomy},
which shows that having a binary symmetric fractional polymorphism is
the \emph{only} reason for tractability, and conversely, that
the condition (MC) is the only reason for intractability.
This provides a complexity
classification of \emph{all} constraint languages defined on \emph{arbitrary} finite
domains, thus generalising all previous classifications mentioned above.

\begin{theorem}[Main]\label{thm:dichotomy}
Let $D$ be an arbitrary finite set, let $\Gamma$ be a 
constraint language defined on $D$, and
let $\Gamma'$ be a core of $\Gamma$.
\begin{itemize}
\item Either $\Gamma$ has a binary symmetric fractional polymorphism and BLP
solves VCSP$(\Gamma)$;
\item or (MC) holds for $\Gamma'_c$ and VCSP$(\Gamma)$ is NP-hard.
\end{itemize}
\end{theorem}

\begin{proof}
If $\Gamma'_c$ satisfies (MC), then
VCSP$(\Gamma'_c)$ is NP-hard by Lemma~\ref{lem:mc}.
In this case VCSP$(\Gamma)$ is NP-hard by Proposition~\ref{prop:core}(\ref{core:hardness})
and Lemma~\ref{lem:core}.
Otherwise, by Theorem~\ref{thm:mainimpl},
$\Gamma'_c$ and hence $\Gamma'$ admit a binary symmetric fractional polymorphism.
By Lemma~\ref{lem:symcore},
$\Gamma$ admits a binary symmetric fractional polymorphism
and it follows from Theorem~\ref{thm:blp} that BLP solves VCSP$(\Gamma)$.
\end{proof}

Theorem~\ref{thm:main} follows immediately from Theorem~\ref{thm:dichotomy}.
We remark that the dichotomy classification 
holds in the special case of $\{0,1\}$-valued
constraint languages, that is, for (weighted) \emph{maximum constraint satisfaction
problems} (Max-CSPs)~\cite{Creignouetal:siam01}.\footnote{We consider Max-CSPs
as Min-CSPs to fit in the VCSP framework; that is, rather than maximising the
(weighted) sum of satisfied constraints the goal is to minimise the (weighted)
sum of unsatisfied constraints. Note that this kind of construction does not
necessarily preserve approximability properties.}

The problem of
deciding whether a constraint language $\Gamma$ is a core and that of deciding
whether the tractability condition of $\Gamma$ is met are discussed in
Section~\ref{sec:meta}.

We discuss constraint languages of \emph{infinite} size in Appendix~\ref{sec:infinite}.

\begin{corollary}[of Theorem~\ref{thm:mainimpl}]
\label{cor}
Let $D$ be an arbitrary finite set and let $\Gamma$ be a core constraint language
defined on $D$. The following are equivalent:
\begin{enumerate}
\item $\Gamma_c$ does not satisfy (MC);
\item $\Gamma$ admits an idempotent and cyclic fractional polymorphism of some arity $k>1$;
\item $\Gamma$ admits an idempotent and symmetric fractional polymorphism of some arity $k>1$;
\item $\Gamma$ admits a binary idempotent and symmetric fractional polymorphism;
\item BLP solves VCSP$(\Gamma)$.
\end{enumerate}
\end{corollary}

\begin{proof}
Theorem~\ref{thm:blp} gives $(4)\Leftrightarrow (5)$.
The implications $(4)\Rightarrow (3)\Rightarrow (2)$ are trivial.
Theorem~\ref{thm:mainimpl} gives the implication $(1)\Rightarrow (4)$.
Finally, we will show that $(2)\Rightarrow (1)$. Let $\omega$ be a $k$-ary
cyclic fractional polymorphism of $\Gamma$. Suppose that $\Gamma_c$
satisfies (MC). By Lemma~\ref{lem:mctomcp}, $\Gamma_c$ satisfies (MC$'$); that is,
there are distinct $a,b\in D$, a unary cost function $u\in\express{\Gamma_c}$ with
$\argmin u=\{a,b\}$, and a binary cost function $f\in\express{\Gamma_c}$ with
$f(a,b)=f(b,a)<f(a,a)=f(b,b)$. 
Consider the
tuples $\tup{a}^1=(a,b)$, $\tup{a}^2=(b,a)$, and $\tup{a}^i=(a,a)$ for $3\leq i\leq k$. 
Note that for every (cyclic) operation $g\in\supp(\omega)$ we have
$g(\tup{a}^1,\ldots,\tup{a}^k)=(x_g,x_g)$ for some $x_g \in D$. 
Using the fact that $\omega$ is a fractional polymorphism of $u$, we first show
that $x_g \in\{a,b\}$.
Observe that 
$\sum_{g} \omega(g) u(g(a^1_1,\ldots,a^k_1)) \leq u^k(a^1_1,\ldots,a^k_1) = 
\frac{k-1}{k}u(a)+\frac{1}{k}u(b)=u(a)=u(b)$,
where the inequality follows from~(\ref{ineq:fpol}).
Hence, we must have $x_g = g(a^1_1,\ldots,a^k_1) \in \{a,b\}$ for all $g \in \supp(\omega)$.
Furthermore, $\sum_{g} \omega(g) f(g(\tup{a}^1, \dots, \tup{a}^k)) = f(a,a)=f(b,b)$,
but $f^k(\tup{a}^1, \dots, \tup{a}^k) = \frac{2}{k}f(a,b)+\frac{k-2}{k}f(a,a) < f(a,a)$.
Thus, inequality~(\ref{ineq:fpol}) does not hold for $f$ and $\omega$.
Consequently, $\omega$ is not a fractional polymorphism of $f$, which is a contradiction.
\end{proof}

Corollary~\ref{cor} answers Problem~1 from~\cite{hkp13:soda} that asked about
the relationship between the complexity of a constraint language $\Gamma$ and
the existence of various types of fractional polymorphisms of $\Gamma$. Note
that Corollary~\ref{cor} holds unconditionally. Problem~1 from~\cite{hkp13:soda}
also involved the solvability by the basic SDP
relaxation~\cite{Raghavendra08:stoc}, which at the time was known to be implied
by (2) and imply (1), provided that P $\neq$ NP. Under the same assumption, we
conclude that solvability by the basic SDP relaxation is also characterised by
any of the equivalent statements of Corollary~\ref{cor}.

\section{Meta problems}\label{sec:meta}

Let $\Gamma$ be a constraint language defined on $D$.
In this section, we study three \emph{meta problems} relevant to our classification.
The first problem is \emph{core recognition}: Given a $\Gamma$, is $\Gamma$ a
core? The second problem is \emph{core identification}: Given $\Gamma$ and 
$\Gamma'$, is $\Gamma'$ a core of $\Gamma$?
The third problem is \emph{tractability recognition}: Given $\Gamma$, is
$\Gamma$ tractable?

We show that all three problems are decidable.
The first two problems are
co-NP-complete and DP-complete, respectively. On the other hand, 
if $\Gamma$ is assumed to be a core, then
the tractability of
$\Gamma$ can be decided in polynomial time.

\begin{lemma}\label{lem:npcontainment}
  Given $\Gamma$ and $g \in \opm{D}{1}$, the problem of deciding whether $\Gamma$
  has a unary fractional polymorphism $\omega$ with $g \in \supp(\omega)$ is in NP.
\end{lemma}

\begin{proof}
  By Lemma~\ref{lem:operationsinomega}, $(\Gamma,g)$ is a yes-instance if, and only if,
  the system of linear inequalities in (\ref{eq:op}) is satisfiable.
  Since the number of inequalities is polynomial in the size of $\Gamma$,
  this system is satisfiable if, and only if, it has a solution with a polynomial number of
  non-zero variables.
  The NP certificate consists of a polynomially large subset of the variables.
  Writing down the restriction of (\ref{eq:op}) to this subset and verifying the satisfiability of
  the resulting system can then be done in polynomial time.
\end{proof}

To every $\{0,1\}$-valued cost function $f$ on domain $D$ corresponds a relation
$R$ defined by $\tup{x} \in R$ if, and only if, $f(\tup{x}) = 0$.
A unary operation $g : D \to D$ is said to be an \emph{endomorphism} of $R$ if
$\tup{x} \in R$ implies $g(\tup{x}) \in R$.

\begin{lemma}\label{lem:endo}
  Let $f$ be a $\{0,1\}$-valued cost function and let $R$ be the corresponding relation.
  The constraint language $\{f\}$ has a unary fractional polymorphism with support
  $\Psi$ if, and only if, $\Psi$ is a set of endomorphisms of $R$.
\end{lemma}

\begin{proof}
  Let $\Psi$ be a set of endomorphisms of $R$ and let $g \in \Psi$, i.e.,
  $\tup{x} \in R$ implies $g(\tup{x}) \in R$, for all $\tup{x} \in D^{ar(f)}$. Then, $f(\tup{x}) \geq f(g(\tup{x}))$, so $\chi_{\{g\}}$ is a unary fractional polymorphism of $\{f\}$.
  It follows that $|\Psi|^{-1} \chi_{\Psi}$ is also a unary fractional polymorphism of $\{f\}$.

  For the opposite direction, let $\omega$ be a unary fractional polymorphism of $\{f\}$. Then,
  
\[
f(\tup{x}) \geq \sum_{g \in \supp(\omega)} \omega(g) f(g(\tup{x})),
\]
for each $\tup{x} \in D^{ar(f)}$.
Fix an operation $g \in \supp(\omega)$.
If $\tup{x} \in R$, then $f(\tup{x}) = 0$ and so clearly $f(g(\tup{x})) = 0$,
i.e., $g(\tup{x}) \in R$.   It follows that $g$ is an endomorphism of $R$.
Since $g \in \supp(\omega)$ was chosen arbitrarily, the result follows.
\end{proof}

\begin{proposition}\label{prop:testing-co-NP-h}
  Testing whether a given constraint language $\Gamma$
  is a core is co-NP-complete.
\end{proposition}

\begin{proof}
  We show that testing whether a given constraint language $\Gamma$ is \emph{not} a core is
  NP-complete.
  Containment in NP follows from Lemma~\ref{lem:npcontainment} by first guessing a non-injective unary operation $g$.
   
  A graph $G$ is a core if all endomorphisms of its edge relation are
  injective~\cite{Hell:graphs}.
  It has been shown in~\cite{Hell92:core} that the problem of checking whether a given
  graph $G$ is not a core is NP-hard, i.e., it is NP-hard to determine whether
  $G$ has a non-injective endomorphism. By Lemma~\ref{lem:endo}, this is the
  case if, and only if, the cost function $f$ corresponding to the adjacency
  relation of $G$ has a unary fractional polymorphism with a non-injective operation in its support, i.e., if, and only if, $\{f\}$ is not a core.
  Therefore, the problem of determining whether $\Gamma$ is not a core is NP-hard, even if $\Gamma$ is only allowed to contain a single binary and symmetric $\{0,1\}$-valued cost function.
\end{proof}

The complexity class DP consists of all decision problems that can be written as the
intersection of an NP-problem and a co-NP-problem; equivalently, DP consists of
all decision problems that can be written as the difference of two
NP-problems~\cite{Papadimitriou84:jcss}.
Next we show that the core identification problem is DP-complete.

\begin{proposition}
  Given two constraint languages $\Gamma$ and $\Gamma'$, testing
  whether $\Gamma'$ is a core of $\Gamma$ is DP-complete.
\end{proposition}
\begin{proof}
  The problem can be described as the intersection between the problem of verifying
  that $\Gamma'$ is a core, which is in co-NP by Proposition~\ref{prop:testing-co-NP-h}, and the
  problem of verifying that $\Gamma' = \Gamma[g(D)]$ for
  some $g$ contained in the support of a unary fractional polymorphism of $\Gamma$.
  The latter problem is seen to be in NP by first guessing the operation $g$, and then
  using Lemma~\ref{lem:npcontainment}.
  Containment in DP follows.

  To show DP-hardness,
  we will reduce from the following problem:
  Given two graphs, $G$ and $G'$, with $G'$ a subgraph of $G$, test
  whether $G'$ is a core (all endomorphisms of $G'$ are injective) and
  whether there is a homomorphism from $G$ to $G'$.
  This problem has been shown to be DP-hard~\cite{Fagin05:core}, thus improving a previously
  known NP-hardness result on the same problem~\cite{Chandra77:stoc}.
  We may in fact assume that $G'$ is an induced subgraph of $G$
  since otherwise, it is easy to see that $G'$ cannot be a core of $G$.
  Let $f$ and $f'$ be the cost functions corresponding to the
  adjacency relations of $G$ and $G'$ respectively.
  Let $\Gamma=\{f\}$ and $\Gamma'=\{f'\}$.
  By Lemma~\ref{lem:endo}, $G'$ is a core
  if, and only if, every unary fractional polymorphism of $\Gamma'$ has only
  injective operations in its support. By Definition~\ref{def:core}, this is the
  case if, and only if, $\Gamma'$ is a core.
  There is a homomorphism from $G$ to $G'$ if, and only if (since $G'$ is a
  subgraph of $G$), there is an endomorphism $g:G\to G$ so that $g(V(G))=V(G')$. By
  Lemma~\ref{lem:endo}, this is the case if, and only if, there is a unary
  fractional polymorphism $\omega$ of $\Gamma$ with $g\in\supp(\omega)$ so that
  $\Gamma'=\Gamma[g(D)]$.
  Hence, $G'$ is a core of $G$ if, and only if, $\Gamma'$ is a core of $\Gamma$.
  It follows that the latter problem is DP-hard,
  even for the specific case when both $\Gamma$ and $\Gamma'$ contains a 
  single binary and symmetric $\{0,1\}$-valued cost function.
\end{proof}

Now we turn our attention to the problem of tractability recognition.
Let $X = \{ (f,\tup{x},\tup{y}) \mid f \in \Gamma, \tup{x}, \tup{y} \in D^{ar(f)} \}$.
To test whether a finite constraint language $\Gamma$ is tractable, it suffices, by
Theorem~\ref{thm:mainimpl}, to test whether it has a binary
symmetric fractional polymorphism. This is the case if, and only if, the
following system of linear inequalities is satisfiable:

\begin{equation}
\begin{aligned}
  \sum_{g\in\Omega} \omega(g) f(g(\tup{x},\tup{y})) & \leq 
  f^2(\tup{x},\tup{y}), \quad \forall (f,\tup{x},\tup{y}) \in X, \\
  \|\omega\|_1 & = 1, \\
  \omega(g) & \geq 0, \quad \forall g\in\Omega,
  \label{eq:expvars}
\end{aligned}
\end{equation}
where $\Omega$ is the set of binary operations $g\in\opm{D}{2}$ on $D$ that are
symmetric.
It follows that the tractability recognition problem is decidable for any finite $\Gamma$.
Since the number of variables in the system (\ref{eq:expvars}) is exponential in
$|D|$, this does not lead to a polynomial-time algorithm.
However, when $\Gamma$ is a core, it turns out that we can solve the system
in polynomial time.
This reflects a well-known phenomenon for the CSP decision problem,
where the problem of deciding whether a constraint language admits various
types of polymorphisms is known to have a polynomial-time algorithm
only when the language is a core.

For a core $\Gamma$, we can restrict $\Omega$ to the set of binary operations on $D$
that are symmetric \emph{and idempotent}.
The linear programming dual of minimising the objective function $0$ subject to~$(\ref{eq:expvars})$
(i.e., of determining whether this system is satisfiable)
is the problem of maximising $\delta$ subject to the following system of inequalities:

\begin{equation}
\begin{aligned}
  \sum_{f, \tup{x}, \tup{y}}  
  z(f,\tup{x},\tup{y}) \left( f^2(\tup{x},\tup{y}) - f(g(\tup{x},\tup{y})) \right) + \delta & \leq 0 ,
  \quad \forall g \in \Omega, \\
  z(f,\tup{x},\tup{y}) & \geq 0, \quad \forall (f, \tup{x}, \tup{y}) \in X.
  \label{eq:expineqs2}
\end{aligned}
\end{equation}

The solution to (\ref{eq:expineqs2}) that assigns $0$ to all variables is always feasible, so
the dual optimum is always at least $0$.
If the dual optimum is $0$,
then the primal optimum is also $0$, so (\ref{eq:expvars}) is satisfiable.
Otherwise, (\ref{eq:expineqs2}) has a solution of measure greater than 0, 
so it has solutions of unbounded measure.
In this case, (\ref{eq:expvars}) is unsatisfiable.
The system (\ref{eq:expineqs2}) has a polynomial number of variables,
but an exponential number of inequalities.

Assuming that $\Gamma$ is a core constraint language,
we can solve (\ref{eq:expineqs2}) in polynomial time using the \emph{ellipsoid method}.
In fact, we can do even better.
We can find a dual solution with support on a polynomial number of variables.
This means that we can find a binary idempotent and symmetric fractional polymorphism
represented by its values on a support of size linear in the size of $X$ and thus in 
the size of $\Gamma$. 
For a thorough treatment of the ellipsoid algorithm, including Lemma~\ref{lem:ellipsoid},
we refer to~\cite{GLS}.

\begin{definition}
  A \emph{strong separation oracle} for a polyhedron $P$ is given
  an input $\tup{p} \in \Q^n$ and either returns
  $``\tup{p} \in P"$, or a vector
  $\tup{a} \in \Q^n$ such that $\tup{a}^\top \tup{x} < \tup{a}^\top \tup{p}$ for all $\tup{x} \in P$.
\end{definition}

\begin{lemma}[Lemma 6.5.15 in~\cite{GLS}]
  \label{lem:ellipsoid}
  Let $\tup{c} \in \Q^n$ and let $P \subseteq \Q^n$ be a polyhedron defined by 
  $A \tup{x} \leq \tup{b}$, where
  the encoding sizes of the coefficients of $A$ and $\tup{b}$ are bounded
  by $\phi$.
  Given a strong separation oracle SEP for $P$ where every output has encoding size at most $\phi$,
  we can, in time polynomial in $n$, $\phi$, 
  and the encoding size of $\tup{c}$, 
  and using a polynomial number of oracle queries to SEP,
  either
  \begin{itemize}
  \item
  find a basic optimum dual solution with oracle inequalities, or
  \item
  assert that the dual problem is unbounded or has no solution.
  \end{itemize}
\end{lemma}

In Lemma~\ref{lem:ellipsoid},
a \emph{basic optimum dual solution with oracle inequalities} means a
set of inequalities
$(\tup{a}^1)^\top \tup{x} \leq \alpha_1, \dots, (\tup{a}^k)^\top \tup{x} \leq \alpha_k$,
valid for $P$,
where $\tup{a}^1, \dots, \tup{a}^k$ are linearly independent outputs of SEP,
and dual variables
$\lambda_1,\dots,\lambda_k \in \Qnn$ such that
$\lambda_1 \tup{a}^1 + \dots + \lambda_k \tup{a}^k = \tup{c}$ and
$\lambda_1 \alpha_1 + \dots + \lambda_k \alpha_k = \max_{\tup{x}\in P} \tup{c}^\top \tup{x}$.

\begin{proposition}
  \label{lem:tractpoly}
  There is a polynomial-time algorithm that,
  given a core constraint language $\Gamma$, either
  \begin{itemize}
  \item 
  finds a binary idempotent and symmetric fractional polymorphism $\omega$ of $\Gamma$,
  represented by a subset $\Omega' \subseteq \Omega$ with $\supp(\omega) \subseteq \Omega'$ together with the restriction of $\omega$ to $\Omega'$, or
  \item
  asserts that none exists.
  \end{itemize}
\end{proposition}

\begin{proof}
  Let $P$ denote the polyhedron defined by (\ref{eq:expineqs2}).
  We will give a polynomial-time algorithm that,
  given a point $(z,\delta) \in \Q^X \times \Q$ as input, does one of three things:
  \begin{itemize}
  \item
    answers ``unbounded optimum''; 
  \item
    answers ``$(z,\delta) \in P$''; or
  \item
    returns $\tup{a} \in \Q^X \times \Q$ such that
  $\tup{a}^\top (x,\delta') < \tup{a}^\top (z,\delta)$ for all $(x,\delta') \in P$.
  \end{itemize}
  The algorithm can be seen as a strong separation oracle with an escape clause.
  We can use it as a strong separation oracle for the polyhedron $P$, as long as the answer is not ``unbounded optimum''.
  
  Let $\tup{c}$ be the vector with components $c_{(f,\tup{x},\tup{y})} = 0$ for $(f, \tup{x}, \tup{y}) \in X$
  and $c_\delta = 1$.
  By Lemma~\ref{lem:ellipsoid}, we can either find a dual solution to (\ref{eq:expineqs2}) given by inequalities returned by the oracle, or we can assert that the dual, (\ref{eq:expvars}), has no solution.
  If the ellipsoid algorithm asserts that the dual has no solution,
  or if the answer from the separation oracle is ever ``unbounded optimum'', then we can conclude that (\ref{eq:expvars}) is unsatisfiable.
  Otherwise, an optimum dual solution is described using valid inequalities of the following form:
  \begin{align*}
  \sum_{(f, \tup{x}, \tup{y}) \in X} z(f,\tup{x},\tup{y}) (f^2(\tup{x},\tup{y})-f(g(\tup{x},\tup{y}))) + \delta &\leq \alpha_g, \qquad \forall g \in \Omega', \\
  -z(f,\tup{x},\tup{y}) &\leq \alpha_{(f,\tup{x},\tup{y})}, \qquad \forall (f,\tup{x},\tup{y}) \in \Upsilon,
  \end{align*}
  for some constants $\alpha_g, \alpha_{(f,\tup{x}, \tup{y})} \in \Q$ and
  subsets $\Omega' \subseteq \Omega$ and $\Upsilon \subseteq X$.
  
  The corresponding dual variables are $\omega' : \Omega' \to \Qnn$ and $\upsilon : \Upsilon \to \Qnn$,
  and they satisfy the following equalities:
  \begin{align}
  \sum_{g \in \Omega'} \omega'(g) (f^2(\tup{x},\tup{y})-f(g(\tup{x},\tup{y}))) - \upsilon(f,\tup{x},\tup{y}) &= 0, \qquad \forall (f,\tup{x},\tup{y}) \in X, \label{eq:dualeq1} \\
  \sum_{g \in \Omega'} \omega'(g) &= 1, \label{eq:dualeq2}
  \end{align}
  where we define $\upsilon(f,\tup{x},\tup{y}) = 0$ for $(f,\tup{x},\tup{y}) \in X \setminus \Upsilon$.
  The dual variables are non-negative, so (\ref{eq:dualeq1}) and (\ref{eq:dualeq2}) imply 
  $f^2(\tup{x},\tup{y}) \geq \sum_{g \in \Omega'} \omega'(g) f(g(\tup{x},\tup{y}))$,
  for all $(f,\tup{x},\tup{y}) \in X$.
  Since the inequalities correspond to vectors that are linearly independent,
  the size of $\Omega'$ is bounded by the number of variables of (\ref{eq:expineqs2}),
  i.e., polynomial in the input size.
  Clearly, $\omega'$ can be extended to a fractional polymorphism of $\Gamma$ by assigning weight 0 to every operation outside of $\Omega'$.

  The separation oracle is given by Algorithm~\ref{alg:separate}.
  It is based on the observation that in order to verify whether
  $(z,\delta)$ belongs to $P$, it suffices to find an operation $g \in \Omega$
  that minimises $\sum_{f, \tup{x}, \tup{y}} z(f,\tup{x},\tup{y}) f(g(\tup{x},\tup{y}))$.
  If $(z,\delta)$ satisfies the inequality with respect to this $g$, then $(z,\delta)$ satisfies all inequalities.
  Otherwise, the vector $\tup{a}$ given by $a_{(f, \tup{x}, \tup{y})} = f^2(\tup{x},\tup{y})-f(g(\tup{x},\tup{y}))$ and $a_{\delta} = 1$ defines a separating hyperplane.

\begin{algorithm}[ht]
\SetAlgoLined
\SetCommentSty{textit}
\DontPrintSemicolon
\SetKwInput{KwPre}{Assumption}
\KwIn{$(z,\delta) \in \Q^X \times \Q$}
\KwOut{``unbounded optimum'', ``$(z,\delta) \in P$'', or a separating hyperplane}
\BlankLine
\If{$z(f,\tup{x},\tup{y}) < 0$ for some $(f,\tup{x},\tup{y}) \in X$}{
  Let $a_{(f,\tup{x},\tup{y})} := -1$ and set all other components of $\tup{a}$ to 0\;
  \Return $\tup{a}$
}\;

Let $V := \{[x,y] \mid x,y \in D\}$ \tcc*[r]{Construct the VCSP instance $I$}
Let $f_I(V) := \sum_{(f,\tup{v}) \in X'}z'(f,\tup{v}) f(\tup{v})$\;

\tcc*[r]{Self reduce using the BLP relaxation}
Let $g' \colon V \to D \cup \{ \bot \}$ be given by $g'(v) = \bot$ for all $v$\;
\While{$\exists v \in V \colon g'(v) = \bot$}{
  \eIf{$\exists d \in D \colon BLP(I[g' \cup \{ v \mapsto d \}]) = BLP(I)$}{
    $g' := g' \cup \{ v \mapsto d \}$
  }{
    \Return ``unbounded optimum''
  }
}

\tcc*[r]{Test whether $(z,\delta) \in P$}
Let $g \in \Omega$ be the operation $(x,y) \mapsto g'([x,y])$\;
\eIf{$\sum_{(f, \tup{x}, \tup{y}) \in X}  z(f,\tup{x},\tup{y}) 
  (f^2(\tup{x},\tup{y}) - f(g(\tup{x},\tup{y}))) + \delta \leq 0$}{
  \Return ``$(z,\delta) \in P$''
}{
  Let $a_{(f, \tup{x}, \tup{y})} := f^2(\tup{x},\tup{y})-f(g(\tup{x},\tup{y}))$, for all $(f,\tup{x},\tup{y}) \in X$,
  and $a_{\delta} := 1$\;
  \Return $\tup{a}$
}
\caption{Separate($z, \delta$)}
\label{alg:separate}
\end{algorithm}

  Let $[x,y]$ denote the multiset of the elements $x$ and $y$,
  and let $V = \{ [x,y] \mid x, y \in D \}$.
  Let $X' = \{ (f,\tup{v}) \mid f \in \Gamma, \tup{v} \in V^{ar(f)} \}$.
  For $(f,\tup{v}) \in X'$, define
  \[
  z'(f,\tup{v}) = \sum_{\substack{\tup{x}, \tup{y} \text{ s.t.}\\ v_i = [x_i,y_i]}} z(f,\tup{x},\tup{y}).
  \]
  The algorithm starts by creating an instance $I$ of VCSP$(\Gamma)$ over the variables $V$ with $f_I(V) = \sum_{(f,\tup{v}) \in X'} z'(f,\tup{v}) f(\tup{v})$.
  For an operation $g \in \Omega$, define the function $g' : V \to D$ by
  $[x,y] \mapsto g(x,y)$.
  Note that this defines a bijection between $\Omega$ and the set of all functions from $V$ to $D$.
  
  For every $g \in \Omega$, we have
  \begin{equation}
    \label{eq:fiv}
    \sum_{(f, \tup{x}, \tup{y}) \in X} z(f,\tup{x},\tup{y}) f(g(\tup{x},\tup{y})) =
    \sum_{(f, \tup{v}) \in X'} \sum_{\substack{\tup{x}, \tup{y} \text{ s.t.}\\ v_i = [x_i,y_i]}} z(f,\tup{x},\tup{y}) f(g(\tup{x},\tup{y}))
    = f_I(g'(V)). 
  \end{equation}
  Instead of optimising the left-hand side of (\ref{eq:fiv}) over all $g \in \Omega$,
  we can optimise $f_I(g'(V))$ over all $g' : V \to D$,
  i.e., we can try to solve the VCSP$(\Gamma)$ instance $I$.
  Note that, since $\Gamma \subseteq \Gamma_c$ (Section~\ref{sec:cores}), 
  $I$ can also be seen as an instance of VCSP$(\Gamma_c)$.
  For a (partial) assignment $g' : V \to D \cup \{ \bot \}$, we
  let $I[g']$ denote the VCSP$(\Gamma_c)$-instance obtained by adding
  the constant unary relations $v = g'(v)$ for $v \in V$ such that $g'(v) \neq \bot$.

  On lines 1--4, the algorithm checks that  all components of $z$ are non-negative. Otherwise, a simple separating hyperplane is returned. 

  On lines 6--7, the algorithm constructs the instance $I$.

  On lines 9--16, it then tries to solve this instance using the BLP relaxation
  and self-reduction.
  This is accomplished by fixing the variables one by one to a value that maintains the BLP optimum (lines 10--12).
  If this succeeds for all variables, then by (\ref{eq:fiv}) and the initial observation,
  we can determine whether the point is contained in $P$ by verifying  a single inequality (line 19).

  Otherwise, the instance $I[g']$ of VCSP$(\Gamma_c)$ has an optimum that is strictly greater than the BLP optimum.
  By Theorem~\ref{thm:blp}, it follows that $\Gamma_c$ does not have a binary symmetric fractional polymorphism.
  Since $\Gamma$ is a core, the same must then be true for $\Gamma$.
  In this case (\ref{eq:expineqs2}) has a non-zero solution, and therefore an unbounded optimum,
  so the algorithm gives the correct answer on line 14.
  
  Finally, we argue that Algorithm~\ref{alg:separate} runs in polynomial time.
  The BLP relaxation of $I$ has size that is polynomial in the size of $z$ and $\Gamma$,
  so the call to BLP$(I[g' \cup \{v \mapsto d\}])$ takes polynomial time.
  The number of calls to BLP is at most $|V| \cdot |D| = \mathcal{O}(|D|^3)$, again polynomial
  in the size of $\Gamma$.
\end{proof}

%
\section{Proof of Theorem~\ref{thm:mainimpl}}

In this section, we prove Theorem~\ref{thm:mainimpl}, which we restate here for the
reader's convenience:

\mainimpl*

\subsection{Proof overview}
\label{sec:overview}

We will need to introduce several important concepts and establish some auxiliary results.
First, using Lemma~\ref{lem:motzkin}, we prove, in Section~\ref{sec:MC}, the following:

\begin{lemma}
\label{lem:exist}
   Let $\Delta$ be an arbitrary constraint language defined on a finite set $D(\Delta)$.
   If $\Delta$ does not satisfy (MC) then $\Delta$ has a binary fractional
   polymorphism $\omega$ such that for each $\{a,b\} \subseteq D(\Delta)$, there exists 
   $g \in \supp(\omega)$ with $\{g(a,b),g(b,a)\} \neq \{a,b\}$.
\end{lemma}

Let $\mathbf{1}$ be the identity mapping in $\opm{D}{m \to m}$.
For a fractional mapping $\sigma$ of arity $m \to m$, let
\[
\mathcal{V}(\sigma) = \{ \mathbf{g}_k \circ \dots \circ \mathbf{g}_1 \circ
\mathbf{1} \mid \mathbf{g}_i \in \supp(\sigma), k \geq 0 \}\,.
\]

Let $G = G(\sigma) = (V(G),E(G))$ be the directed graph with
\begin{itemize}
\item
  $V(G) = \mathcal{V}(\sigma)$; 
\item
  $E(G) = \{(\mathbf{g},\mathbf{h} \circ \mathbf{g}) \mid \mathbf{g} \in \mathcal{V}(\sigma), \mathbf{h} \in \supp(\sigma) \}$.
\end{itemize}

A vertex $\mathbf{g}$ in $V(G)$ is called \emph{recurrent} if, for every other
vertex $\mathbf{h} \in V(G)$, there is a path from $\mathbf{h}$ to $\mathbf{g}$
whenever there is a path from $\mathbf{g}$ to $\mathbf{h}$.
Let $\mathcal{R}(\sigma)$ denote the set of maximal strongly connected components of
recurrent vertices of $V(G)$.
Note that $\mathcal{R}(\sigma)$ is a partition of the set of recurrent vertices.

If $\rho$ is a generalised fractional polymorphism of a cost function $f$, then we say that
$\rho$ improves $f$.
The set of all cost functions that are improved by $\rho$ is denoted by $\imp(\rho)$.
The following result is proved in Section~\ref{sec:proofs}.

\begin{theorem}\label{thm:master}
   Let $\sigma$ be a fractional mapping of arity $m \to m$.
   There exists a probability distribution $w$ on $\mathcal{R}(\sigma)$ with the following property:
   if $\rho$ is any fractional mapping of arity $m \to m$ with
   $\sum_{\mathbf{g} \in C} \rho(\mathbf{g}) = w(C)$ for all $C \in \mathcal{R}(\sigma)$,
   then $\imp(\sigma) \subseteq \imp(\rho)$.
\end{theorem}

As the first step in our proof of Theorem~\ref{thm:mainimpl}, 
we apply Lemma~\ref{lem:exist} to $\Gamma_c$.
By assumption, $\Gamma_c$ does not satisfy (MC), so we conclude that it has a fractional
polymorphism $\omegaour$ with the properties given in the lemma. Furthermore, by
Proposition~\ref{prop:core}(1), we know that $\express{\Gamma_c}$
contains a unary function $u_a$ for each $a \in D$ such that $\argmin u_a =
\{a\}$. This implies that $\omegaour$ is idempotent. To finish the proof, we will
massage $\omegaour$ into a binary symmetric fractional polymorphism
using Theorem~\ref{thm:master}.

For a binary operation $g \in \opm{D}{2}$, define $\sym{g}$ by $\sym{g}(x,y) =
g(y,x)$.
We denote by $(g,\sym{g}) \in \opm{D}{2 \to 2}$ the mapping defined by
$(g,\sym{g})(x,y)=(g(x,y),\sym{g}(x,y))$.
Recall that $\chi_{(g,\sym{g})}$ denotes the fractional mapping that that takes
the value 1 on the mapping $(g,\sym{g})$ and 0 on all other mappings. 
Let $\sigmaour = \sum_{g} \omegaour(g) \chi_{(g,\sym{g})}$. 
As the second step, we apply Theorem~\ref{thm:master} to $\sigmaour$.
Note that $\Gamma_c \subseteq \imp(\sigmaour)$ and that all $\mathbf{g} \in \mathcal{V}(\sigmaour)$ are of the form $\mathbf{g} = (g,\sym{g})$.
Let $w$ be the probability distribution in Theorem~\ref{thm:master} when applied to $\sigmaour$.
Fix an arbitrary mapping $\mathbf{g}_C \in C$, for every $C \in \mathcal{R}(\sigmaour)$,
and let $\rhoour = \sum_C w(C) \chi_{\mathbf{g}_C}$.

A mapping $\mathbf{p} \in \opm{D}{m\to m}$ is called \emph{permuting} if it acts
as a permutation on every tuple in $D^m$.
The following lemma finishes the proof of Theorem~\ref{thm:mainimpl}:

\begin{lemma}[Key lemma]\label{lem:main}
  For every $f \in \imp(\rhoour)$, $\tup{x}^1,\tup{x}^2 \in D^{ar(f)}$, 
  $\mathbf{g} \in \supp(\rhoour)$, and 
  permuting mapping $\mathbf{p} \in \opm{D}{2\to 2}$, we have
  $f^2(\mathbf{g}(\tup{x}^1,\tup{x}^2)) = f^2(\mathbf{g} \circ \mathbf{p}(\tup{x}^1,\tup{x}^2))$.
\end{lemma}

\begin{corollary}\label{cor:order}
  For every permuting mapping $\mathbf{p} \in \opm{D}{2\to 2}$, we have
  $\imp(\rhoour) \subseteq \imp(\rhoour \circ \mathbf{p})$,
  where $\rhoour \circ \mathbf{p} := \sum_{\mathbf{g} \in \supp(\rhoour)} \rho(\mathbf{g}) \chi_{\mathbf{g} \circ \mathbf{p}}$.
\end{corollary}

Let $\mathbf{p} \in \opm{D}{2 \to 2}$ be a mapping that orders its inputs
according to some fixed total order on $D$. 
By Theorem~\ref{thm:master} and Corollary~\ref{cor:order}, we have
$$\Gamma \subseteq \Gamma_c \subseteq \imp(\sigmaour) \subseteq \imp(\rhoour) \subseteq \imp(\rhoour \circ \mathbf{p}),$$
so $\Gamma$ admits $\rhoour \circ \mathbf{p}$.
For every $a, b \in D$,
$\mathbf{p}(a,b) = \mathbf{p}(b,a)$
so for every $\mathbf{g} \in \supp(\rhoour)$, we have
$\mathbf{g} \circ \mathbf{p}(a,b) = \mathbf{g} \circ \mathbf{p}(b, a)$.
It follows that $\rhoour \circ \mathbf{p}$ is symmetric.
Consequently,
\[
\sum_{(g_1,g_2) \in \supp(\rhoour \circ \mathbf{p})} \rhoour \circ \mathbf{p}((g_1,g_2))\frac{1}{2}(\chi_{g_1}+\chi_{g_2})
\]
is a binary idempotent and symmetric fractional polymorphism
of $\Gamma$ which proves Theorem~\ref{thm:mainimpl}.

It remains to prove Lemma~\ref{lem:main}.
For this we need two additional results that are stated here and are proved in
Sections~\ref{sec:connected} and~\ref{sec:proofs2}.

\begin{definition}
  Let $w_a = \sum_{\mathbf{g} : \mathbf{g}(a,b) = (a,a)} \rhoour(\mathbf{g})$
  and $w_b = \sum_{\mathbf{g} : \mathbf{g}(a,b) = (b,b)} \rhoour(\mathbf{g})$.
  We say that $\rhoour$ is \emph{submodular on the pair} $\{a,b\} \subseteq D$ if
  $w_a = w_b = \frac{1}{2}$.
\end{definition}

Let $S = (V(S),E(S))$ be the undirected graph with:
\begin{itemize}
\item
  $V(S) = D$;
\item
  $E(S) = \{ \{a,b\} \mid \rhoour$ is submodular on $\{a,b\} \}$.
\end{itemize}

\begin{lemma}\label{lem:connected}
  The graph $S$ is connected.
\end{lemma}

\begin{lemma}\label{lem:mix}
  Assume that $\rhoour$ is submodular on $\{a_1,a_2\}$.
  Let $f \in \imp(\rhoour)$
  and $(\tup{y}^1,\tup{y}^2) = \mathbf{g}(\tup{x}^1,\tup{x}^2)$ for some
  $\mathbf{g} \in \supp(\rhoour)$ and $\tup{x}^1,\tup{x}^2 \in D^{ar(f)-1}$.
  Then
  $f^2((a_1,\tup{y}^1),(a_2,\tup{y}^2)) = f^2((a_2,\tup{y}^1),(a_1,\tup{y}^2))$.
\end{lemma}

\begin{proof}[Proof of Lemma~\ref{lem:main}]
  By construction, 
  $(g, \sym{g})(y,x) = (\sym{g}, g)(x,y)$ for all $(g,\sym{g}) \in \mathcal{V}(\sigmaour)$.
  Therefore, it suffices to show that interchanging the two elements of
  $\mathbf{g}(\tup{x}^1,\tup{x}^2)$ at any subset of the coordinates 
  does not alter the value of $f^2(\mathbf{g}(\tup{x}^1,\tup{x}^2))$.
  We show this for the case when
  only the elements of the first coordinate are interchanged:
  with $\mathbf{g}(\tup{x}^1,\tup{x}^2) = ((a,\tup{y}^1),(b,\tup{y}^2))$,
  we show that $f^2((a,\tup{y}^1),(b,\tup{y}^2)) =
  f^2((b,\tup{y}^1),(a,\tup{y}^2))$.
  The full result follows by applying the same argument to each coordinate.
  By Lemma~\ref{lem:connected}, there exists a path
  $a = a_0, a_1, \dots, a_\ell = b$ from $a$ to $b$ in the graph $S$,
  and by Lemma~\ref{lem:mix}, we have
  \begin{equation}\label{eq:oneflip}
    f^2((a_i,\tup{y}^1),(a_{i+1},\tup{y}^2)) =
    f^2((a_{i+1},\tup{y}^1),(a_i,\tup{y}^2)),
  \end{equation}
  for all $0 \leq i < \ell$.
  Summing (\ref{eq:oneflip}) over $0 \leq i < \ell$, we obtain
  \begin{equation}\label{eq:sumflip}
    \sum_{0 \leq i < \ell} f^2((a_i,\tup{y}^1),(a_{i+1},\tup{y}^2)) = 
    \sum_{0 \leq i < \ell} f^2((a_{i+1},\tup{y}^1),(a_i,\tup{y}^2)).
  \end{equation}
  
  Finally, by cancelling terms in (\ref{eq:sumflip}),
  \[
  \frac{1}{2} f((a_0,\tup{y}^1)) + \frac{1}{2} f((a_\ell,\tup{y}^2)) =
  \frac{1}{2} f((a_\ell,\tup{y}^1)) + \frac{1}{2} f((a_0,\tup{y}^2)),
  \]
  which establishes the result.
\end{proof}

\subsection{Proof of Lemma~\ref{lem:symcore}}
\label{sec:newlemmaproof}
Here, we use Theorem~\ref{thm:master} to prove Lemma~\ref{lem:symcore}.

\begin{proof}[Proof of Lemma~\ref{lem:symcore}]
Let $\omega'$ be a binary symmetric fractional polymorphism of $\Gamma'$.
Let $D' \subseteq D$ be the domain of the core $\Gamma'$,
and 
let $\mu$ be a unary fractional polymorphism of $\Gamma$ with $g \in \supp(\mu)$ such that
$\Gamma' = \Gamma[g(D)]$ and thus $D'=g(D)$.
Consider the graph $G(\mu)$,
and define the fractional operation $\mu'$
as follows:
for each component $C \in \mathcal{R}(\mu)$, pick any unary operation $h \in C$, note that
$g \circ h \in C$, and let
$\mu'(g \circ h) = w(C)$.
Then, by Theorem~\ref{thm:master},
$\mu'$ is a unary fractional polymorphism of $\Gamma$ with the property that
$h'(D) \subseteq g(D) = D'$ for every $h' \in \supp(\mu')$.

Now define the following fractional operation:
\[
\omega := \sum_{g' \in \supp(\omega')}
\omega'(g')
\sum_{h' \in \supp(\mu')}
\mu'(h')\chi_{g \circ (h', h')}.
\]

Let $f \in \Gamma$ and $\tup{x}^1, \tup{x}^2 \in D^{ar(f)}$.
Then,
\begin{align*}
f^2(\tup{x}^1, \tup{x}^2) &\leq
\sum_{h' \in \supp(\mu')} \mu'(h') f^2(h'(\tup{x}^1), h'(\tup{x}^2)) \\
&\leq
\sum_{h' \in \supp(\mu')}
\mu'(h')
\sum_{g' \in \supp(\omega')}
\omega'(g')
f^2(g'(h'(\tup{x}^1), h'(\tup{x}^2))) \\
&=
\sum_{g \in \supp(\omega)} \omega(g) f^2(g(\tup{x}^1, \tup{x}^2)),
\end{align*}
where the first inequality follows since $\Gamma$ admits $\mu$,
and the second inequality follows since $\Gamma'$ admits $\omega$,
and $h'(\tup{x}^1), h'(\tup{x}^2) \in (D')^{ar(f)}$ for every $h' \in \supp(\mu')$.
Hence, $\omega$ is a binary symmetric fractional polymorphism of $\Gamma$,
which proves the lemma.
\end{proof}

\subsection{Proof of Lemma~\ref{lem:exist}}
\label{sec:MC}

We remark that after the announcement of our work in~\cite{tz13:stoc}, the idea
in the following proof has been used to prove a generalisation of Lemma~\ref{lem:exist}
in~\cite[Lemma~28]{Uppman14:stacs}, where it has been used to analyse the
complexity of certain Min-Cost-Hom problems.

\begin{proof}[Proof of Lemma~\ref{lem:exist}]
  Let $\pi_1(x,y) = x$ and $\pi_2(x,y) = y$ be the two binary projections on $D(\Delta)$.
  Let $\Omega(a,b)$ be the set of operations $g : D(\Delta) \times D(\Delta) \to D(\Delta)$ 
  for which $\{g(a,b),g(b,a)\} \neq \{a,b\}$.
  Assume that there exist rational values $y(f,\tup{x}) \geq 0$,
  for
  $f \in \Delta, \tup{x} \in (D(\Delta) \times D(\Delta))^{ar(f)}$, such that
  \begin{align}
    \sum_{f,\tup{x}} y(f,\tup{x}) f(g(\tup{x})) & \geq \sum_{f,\tup{x}}
    y(f,\tup{x}) f(\pi_i(\tup{x})), \quad \forall g \in \opm{D(\Delta)}{2}, i = 1, 2, \label{eqn:maxcut1} \\
    \sum_{f,\tup{x}} y(f,\tup{x}) f(g(\tup{x})) & > \sum_{f,\tup{x}}
    y(f,\tup{x}) f(\pi_i(\tup{x})), \quad \forall g \in \Omega(a,b), i = 1, 2. \label{eqn:maxcut2}
  \end{align}
  
  Let $V = \{ v_{(x,y)} \mid (x,y) \in D(\Delta) \times D(\Delta) \}$ and
  let $v_1,\dots,v_n$ be an enumeration of $V$ with $v_1 = v_{(a,b)}$ and $v_2 = v_{(b,a)}$.
  Define $\iota : V \to D(\Delta) \times D(\Delta)$ by $\iota(v_{(x,y)}) = (x,y)$
  and
  let $I$ be the instance of VCSP$(\Delta)$ with variables $V$ and
  objective function $f_I(v_1, \dots, v_n) = \sum_{f, \tup{x}} y(f,\tup{x}) f(\iota^{-1}(\tup{x}))$.
  Define $f(x,y) = \min_{v_3,\dots,v_n \in D} f_I(x,y,v_3,\dots,v_n) \in \express{\Delta}$.
  The equations (\ref{eqn:maxcut1}) imply that $\pi_1 \circ \iota$ and $\pi_2 \circ \iota$ are among the
  optimal solutions to $I$, and
  the equations (\ref{eqn:maxcut2}) imply that $\pi_1 \circ \iota$ and $\pi_2 \circ \iota$ have strictly
  smaller measure than any solution $g \in \Omega(a,b)$,
  so $f(a,b) = f(b,a) < f(x,y)$ for all $\{x,y\} \neq \{a,b\}$.
  
  We conclude that if (MC) cannot be satisfied,
  then
  there is no solution to the system (\ref{eqn:maxcut1})+(\ref{eqn:maxcut2}).
  By Lemma~\ref{lem:motzkin}, there is a solution $z_1(g,i), z_2(g,i) \geq 0$ to the following system of equations:

  \begin{equation}
  \begin{aligned}
  \label{eqn:fpol}
    & \sum_{i=1}^2 \sum_{g \in \Omega(a,b)} z_1(g,i) (f(g(\tup{x})) - f(\pi_i(\tup{x}))) \\
    + & \sum_{i=1}^2 \sum_{g \in \opm{D(\Delta)}{2}}z_2(g,i) (f(g(\tup{x}))
    - f(\pi_i(\tup{x})))  \leq 0, \quad
    \forall f \in \Delta, \tup{x} \in (D(\Delta) \times D(\Delta))^{ar(f)},
  \end{aligned}
  \end{equation}
  with $z_1(g,i) \neq 0$ for some $g \in \Omega(a,b)$ and $i \in \{1,2\}$.
  Define $z_1(g,i) = 0$ for $g \not\in \Omega(a,b)$ and let
  $z(g) =  \| z_1 + z_2 \|^{-1} (z_1(g,1)+z_1(g,2)+z_2(g,1)+z_2(g,2))$.
  A solution to (\ref{eqn:fpol}) then implies a solution to the following system of inequalities:

  \begin{equation*}\label{eqn:norm}
    \sum_{g \in \opm{D(\Delta)}{2}} z(g) f(g(\tup{x}))
    \leq f^2(\pi_1(\tup{x}),\pi_2(\tup{x})),
    \quad \forall f \in \Delta, \tup{x} \in (D(\Delta) \times D(\Delta))^{ar(f)},
  \end{equation*}
  with $\|z\|_1 = 1$, $z(g) \geq 0$, and $z(g) > 0$ for some $g \in \Omega(a,b)$.
  Denote this solution by $z_{a,b}(g)$.
  Now, if (MC) cannot be satisfied for \emph{any} distinct $a, b \in D(\Delta)$, then
  we have solutions $z_{a,b}(g)$ for all $a \neq b \in D(\Delta)$.
  The lemma follows with $\omega$ defined by $\omega(g) = (|D(\Delta)|^2-|D(\Delta)|)^{-1} \sum_{a \neq b} z_{a,b}(g)$.
\end{proof}

\subsection{Proof of Lemma~\ref{lem:connected}}
\label{sec:connected}

The aim of this section is to prove that the graph $S$ of submodular pairs is connected.
In order to do so, we introduce yet another graph $T$
that records the ``definable 2-subsets of $D$ in $\express{\Gamma_c}$''.
We then show that $T$ is a subgraph of $S$ and that $T$ is connected.
Since $S$ and $T$ are defined on the same set of vertices, it then follows that $S$ is connected.

Let $T = (V(T),E(T))$ be the undirected graph with:
\begin{itemize}
\item
  $V(T) = D$;
\item
  $E(T) = \{ \{a,b\} \mid$ there exists a unary function  
  $u \in \express{\Gamma_c}$ such that $\argmin u = \{a,b\} \}$.
\end{itemize}

\begin{lemma}
  $E(T) \subseteq E(S)$.
\end{lemma}

\begin{proof}
  Take an arbitrary edge $\{a,b\} \in E(T)$
  and
  let $u_a$, $u_b$, and $u_{ab}$ be unary cost functions in $\express{\Gamma_c}$ such that $\argmin u_a = \{a\}$, $\argmin u_b = \{b\}$, and $\argmin u_{ab} = \{a,b\}$,
  respectively.
  Since $u_{ab}$ minimises on $\{a,b\}$ and is improved by both $\omegaour$ and $\rhoour$,
  we have $g(a,b),g(b,a) \in \{a,b\}$ for every $g \in \supp(\omegaour)$ and 
  every $\mathbf{g} = (g,\bar{g}) \in \supp(\rhoour)$.
  By construction of $\sigmaour$, there is a mapping $\mathbf{h} \in \supp(\sigmaour)$ for which $\mathbf{h}(a,b) \not\in \{(a,b),(b,a)\}$,
  so by our previous observation,
  we must have either $\mathbf{h}(a,b) = (a,a)$ or $\mathbf{h}(a,b) = (b,b)$.
  Suppose that $\mathbf{g}(a,b) \in \{(a,b),(b,a)\}$ for some $\mathbf{g} \in \supp(\rhoour)$.
  Then $\mathbf{h} \circ \mathbf{g}(a,b) = (a,a)$ or $(b,b)$.
  So $\mathbf{h} \circ \mathbf{g}$ is reachable from $\mathbf{g}$ in $G$,
  it is symmetric on $\{a,b\}$,
  and every $\mathbf{g}'$ reachable from $\mathbf{h} \circ \mathbf{g}$ is symmetric on $\{a,b\}$.
  Therefore $\mathbf{g}$ cannot be recurrent.
  But $\supp(\rhoour)$ is contained in the set of recurrent states,
  a contradiction.
  We conclude that every $\mathbf{g} \in \supp(\rhoour)$ is symmetric on $\{a,b\}$
  and maps $(a,b)$ to either $(a,a)$ or $(b,b)$.

  Let $w_a = \sum_{\mathbf{g} : \mathbf{g}(a,b) = (a,a)} \rho(\mathbf{g})$ and
  $w_b = \sum_{\mathbf{g} : \mathbf{g}(a,b) = (b,b)} \rhoour(\mathbf{g})$.
  By the previous argument, we have $w_a + w_b = 1$.
  By the fractional polymorphism inequality applied to $\rhoour$ and $u_{a}$, we have
  \begin{equation}\label{eq:ua}
    \frac{1}{2} (u_{a}(a)+u_{a}(b)) \geq w_a u_{a}(a) + w_b u_{a}(b).
  \end{equation}

  Since $u_{a}(a) < u_{a}(b)$, we have $w_a \geq w_b$.
  But inequality (\ref{eq:ua}) holds for $u_{b}$ as well,
  hence $w_a \leq w_b$, and therefore $w_a = w_b = \frac{1}{2}$.
\end{proof}

\begin{lemma}\label{lemma:Tconnected}
  $T$ is connected.
\end{lemma}

To prove this lemma, we will introduce some terminology from the study of hyperplane arrangements which will facilitate our reasoning about the edges of $T$.
For a more thorough treatment of this subject, see~\cite{buildings} and~\cite{hpintro}.

\begin{definition}
  Let $\{ \tup{v}^i \}_{i \in I}$ be a finite set of vectors in $\R^n$.
  The set of hyperplanes $\mathcal{A} = \{ H_i \}_{i \in I}$, where
  $H_i = \{ \tup{x} \in \R^n \mid \tup{v}^i \cdot \tup{x} = 0 \}$,
  is called a \emph{(linear) hyperplane arrangement}.
\end{definition}

To each vector $\tup{x} \in \R^n$, we associate a \emph{sign vector},
$\sign(\tup{x}) \in \{-1,0,+1\}^I$, where the $i$th component is given by the sign of
$\tup{v}^i \cdot \tup{x}$ for each $i \in I$.
For a sign vector $\tup{v} \in \{-1,0,+1\}^I$, 
a non-empty set
$A = \sign^{-1}(\tup{v}) = \{ \tup{x} \in \R^n \mid \sign(\tup{x}) = \tup{v} \}$ is
called a \emph{cell} of $\mathcal{A}$.
We denote the defining sign vector, $\tup{v}$ of $A$, by $\sign(A)$.

A cell $\ch$ with $\sign(\ch)_i \neq 0$ for all $i \in I$ is called a \emph{chamber}.
The chambers are the connected full-dimensional regions of $\R^n \setminus \bigcup_{i \in I} H_i$.
A cell $P$ with $\sign(P)_i = 0$ for exactly one $i \in I$ is called a \emph{panel}.
We say that \emph{$P$ is a panel of a chamber $\ch$} if the panel $P$ is contained in the topological closure ${\rm cl}(\ch)$ of $\ch$.
Each panel is a panel of precisely two chambers.

The \emph{chamber graph} of $\mathcal{A}$ is the undirected graph with the chambers of $\mathcal{A}$ as vertices and an edge between two chambers $\ch_1$ and $\ch_2$ if $\sign(\ch_1)$ and $\sign(\ch_2)$ differ by a single sign change, or equivalently, if $\ch_1$ and $\ch_2$ share a common panel.
We will use the following properties of the chamber graph that can be found
in~\cite[Proposition~1.54]{buildings}.

\begin{proposition}
\label{prop:gconnected}
  The chamber graph of $\mathcal{A}$ is connected and
  the minimal length of a
  path between $\ch_1$ and $\ch_2$ in the chamber graph
  is equal to the number of positions at which $\sign(\ch_1)$ and $\sign(\ch_2)$ differ.
\end{proposition}

We are now ready to prove Lemma~\ref{lemma:Tconnected}.

\begin{proof}[Proof of Lemma~\ref{lemma:Tconnected}]
  For each $a \in D$, we have a unary function $u_a \in \express{\Gamma_c}$
  with $\argmin u_a = \{a\}$.
  For $\tup{x} \in \R^{D}$, with components $x_c$,
  consider the linear combination $f_{\tup{x}}(z) = \sum_{c \in D} x_c u_c(z)$.
  Note that if $\tup{x}$ is rational and nonnegative,
  then $f_{\tup{x}} \in \express{\Gamma_c}$.
  The inequality
  $f_{\tup{x}}(a) < f_{\tup{x}}(b)$ is equivalent to
  $\sum_{c \in D} x_c (u_c(a)-u_c(b)) < 0$, i.e.,
  $f_{\tup{x}}$ takes a strictly smaller value on $a$ than on $b$ precisely
  when the vector $\tup{x}$ is on the negative side of the hyperplane 
  $H^{ab}$ defined by the normal $\tup{v}^{ab}$ with components $v^{ab}_c = u_c(a)-u_c(b)$.
  Hence, by determining the sign of $\tup{x} \cdot \tup{v}^{ab}$,
  we can decide whether $f_{\tup{x}}(a) < f_{\tup{x}}(b)$ or $f_{\tup{x}}(a) >
  f_{\tup{x}}(b)$.
  If $\tup{x}$ lies \emph{on} the hyperplane, then $f_{\tup{x}}(a) = f_{\tup{x}}(b)$.

  For each $a \in D$, let $H^{a}$ be the hyperplane defined by the \emph{unit
  vector} $\tup{e}^a$, i.e., $e^a_a = 1$ and $e^a_c = 0$ for $a \neq c$.
  Fix a strict total order $<_{D}$ on $D$.
  Let $\mathcal{A} = \{ H^{ab} \mid a <_{D} b \} \cup \{ H^a \mid a \in D \}$
  be a hyperplane arrangement in $\R^{D}$.
  Let $\mathcal{C}$ be the set of chambers $\ch$ that have a positive sign for each $H^a$,
  i.e., each $\ch \in \mathcal{C}$ is contained in the positive (open) orthant of $\R^{D}$.
  Since all remaining components of $\ch \in \mathcal{C}$ are also nonzero, 
  they determine a strict order on the values of the functions 
  $f_{\tup{x}}$, $\tup{x} \in \ch$.
  For each $a \in D$, let
  $U_a = \{ \ch \in \mathcal{C} \mid \forall \tup{x} \in \ch : \argmin f_{\tup{x}} = \{a\} \}$.
  Each $U_a$ is non-empty since the vector $\tup{x}$ given by $x_c = \epsilon$ for $c \neq a$ and $x_a = 1$ determines a function minimizing on $a$ when $\epsilon > 0$ is chosen small enough.

  Fix $a, b \in D$ and pick any $\ch_a \in U_a, \ch_b \in U_b$.
  Let $\ch_a = \ch_0, \ch_1, \dots, \ch_\ell = \ch_b$ be a minimal-length path from $\ch_a$ to $\ch_b$
  in the chamber graph.
  Consider the sign vectors along this path:
  $\sign(\ch_0), \sign(\ch_1), \dots, \sign(\ch_\ell)$.
  By Proposition~\ref{prop:gconnected} the sign of a fixed component changes at most 
  once along this sequence.
  In particular, since $\ch_a$ and $\ch_b$ both have positive signs for the hyperplanes $H^a$, it follows that $\ch_i$ is contained in the positive orthant for every $i$.
  Hence, for each $i$, there is a $a_i \in D$ such that $\ch_i \in U_{a_i}$.
  For each $i$ with $a_i \neq a_{i+1}$, the path moves from
  a chamber where $f_{\tup{x}}$ minimises on $a_i$ to a chamber where it minimises on $a_{i+1}$.
  This means that $\ch_i$ and $\ch_{i+1}$ share a panel $P_i$ with a sign vector $\sign(P_i)$
  obtained from either $\sign(\ch_i)$ or $\sign(\ch_{i+1})$ by setting the component corresponding to $H^{a_ia_{i+1}}$ to 0 (assuming $a_i <_{D} a_{i+1}$).
  Since all other components of $\sign(P_i)$ have the same sign as in
  $\sign(\ch_i)$ and $\sign(\ch_{i+1})$, we have $f_{\tup{x}}(a_i) =
  f_{\tup{x}}(a_{i+1}) < f_{\tup{x}}(c)$, for every $\tup{x} \in P_i$ and $c \neq a_i,a_{i+1}$.
  For a hyperplane arrangement, such as $\mathcal{A}$, that is defined in terms of rational normal vectors, each cell is defined as the solutions to a set of linear equalities and inequalities with rational coefficients. Every cell therefore contains at least one rational vector.
  In particular, there exists a nonnegative rational vector $\tup{x} \in P_i$ with
  $\argmin f_{\tup{x}} = \{a_i,a_{i+1}\}$, so $\{a_i,a_{i+1}\} \in E(T)$.
  This holds for all $0 \leq i < \ell$ with $a_i \neq a_{i+1}$, so we conclude
  that a subsequence of $a = a_0, a_1, \dots, a_\ell = b$ is a path in $T$ from $a$ to $b$.
\end{proof}

\subsection{Proof of Theorem~\ref{thm:master}} 
\label{sec:proofs}

A (time-homogeneous) finite-state Markov chain $M$ is given by a set of states
and conditional probabilities $p(i,j)$ for $M$ to be in state $j$ at time
$t+1$ given that it was in state $i$ at time $t$.
Let $p^{(k)}(i,j)$ denote the probability that $M$ proceeds from state $i$
to state $j$ in exactly $k$ transitions.
$M$ is called \emph{irreducible} if, for every pair of states
$(i,j)$, there exists $r \geq 1$ with $p^{(r)}(i,j) > 0$.
A state $i$ is called \emph{transient} if, for some state $j$, there is a path
(in the graph whose vertices are the states of $M$ and with and edge $(i,j)$
from state $i$ to state $j$ if $p(i,j)>0$) from
$i$ to $j$ but not from $j$ to $i$.
A state that is not transient is called \emph{recurrent}.
A state $i$ has \emph{periodicity} $r$ if $r = \gcd \{ k \mid p^{(k)}(i,i) > 0 \}$.
$M$ is called \emph{aperiodic} if all states have periodicity 1.
A \emph{stationary distribution} of $M$ is a probability distribution $\lambda$
on the set of states of $M$ such that
$\lambda(i) = \sum_{j} \lambda(j) p(j,i)$ for all states $i$.
The following is well known.

\begin{theorem}\label{thm:markov}

    For any finite-state Markov chain $M$:

  \begin{enumerate}
  \item
    If $M$ is irreducible,
    then there is a unique stationary distribution
    $\lambda$ of $M$ with $\lambda(i) > 0$ for all states $i$.
  \item
    If $M$ is aperiodic,
    then for any initial distribution $\pi$,
    there is a stationary distribution $\lambda$ of $M$ with
    $\sum_{j} \pi(j) p^{(k)}(j,i) \to \lambda(i)$ as $k \to \infty$,
    for all states $i$.
  \item
    If $i$ is transient, then $p^{(k)}(j,i) \to 0$ as $k \to \infty$,
    for all states $j$.
  \end{enumerate}
\end{theorem}

\begin{proof}
Part (1) follows from~\cite[Theorem~5.1.1 and~5.1.2]{Kemeny76:finite},
where an irreducible chain is called ergodic.
(The definition in~\cite{Kemeny76:finite} of an ergodic chain differs from the more
common one which defines an ergodic chain as an irreducible and aperiodic chain.)

\cite[Theorem~4.1.4]{Kemeny76:finite}
proves the claim of part (2) for irreducible aperiodic chains.
This result can be extended to any aperiodic chain by considering what happens
for an initial distribution concentrated on a single state $i$.
Let $\mathcal{R}$ denote the set of maximal strongly connected components in
the directed graph that has the recurrent states of $M$ as vertices and
an edge from $i$ to $j$ if $p(i,j) > 0$.
If $i$ is recurrent, then the restriction of $M$ to the component $C \in \mathcal{R}$
containing $i$ is irreducible, so the chain converges to a stationary distribution on $C$
with the desired properties.
Let $\lambda^i$ be the trivial extension of this distribution to a stationary distribution on $M$.
If instead $i$ is transient, then for each component $C \in \mathcal{R}$,
there is some probability that $i$ reaches $C$.
The stationary distribution $\lambda^i$ is then defined as the unique stationary distribution of each
irreducible component, weighted by the probability that $i$ reaches this component.
Finally, the full statement of part (2) follows by taking $\lambda = \sum_i \pi(i) \lambda^i$.

Part (3) follows from~\cite[Theorem~3.1.1]{Kemeny76:finite}.
\end{proof}

Given an $m \to m$ fractional mapping $\sigma$,
we define a Markov chain $M(\sigma)$ on $G(\sigma)$.
Let $w(\mathbf{g},\mathbf{g}') = \sum_{\mathbf{h} \in \supp(\sigma) : \mathbf{g}' = \mathbf{h} \circ \mathbf{g}} \sigma(\mathbf{h})$.
The transition probabilities are given as follows:
\[
p(\mathbf{g},\mathbf{g}') = \begin{cases}
  \frac{1}{2} w(\mathbf{g},\mathbf{g}') + \frac{1}{2} & \text{if $\mathbf{g} = \mathbf{g}'$, and}\\
  \frac{1}{2} w(\mathbf{g},\mathbf{g}') & \text{otherwise.}
\end{cases}
\]

Note that the set of recurrent vertices in $\mathcal{V}(\sigma)$, defined in
Section~\ref{sec:overview}, is precisely the set of
recurrent states of $M(\sigma)$.
Let $C$ be a component in $\mathcal{R}(\sigma)$.
Define $M(C)$ to be the restriction of $M(\sigma)$ to $C \subseteq \mathcal{V}(\sigma)$.
Then, $M(C)$ is also a Markov chain.

\begin{lemma}\label{lem:M}
  The Markov chains $M(\sigma)$ and $M(C)$ are aperiodic and each chain $M(C)$ is irreducible.
\end{lemma}

\begin{proof}
  Aperiodicity follows by construction as $p(\mathbf{g},\mathbf{g}) \geq \frac{1}{2} > 0$ for all $\mathbf{g} \in \mathcal{V}(\sigma)$.
  Irreducibility follows since each $C$ is a maximal strongly connected component of recurrent states.
\end{proof}

\begin{lemma}\label{lem:distr}
  Let $\rho$ and $\lambda$ be probability distributions on $\mathcal{V}(\sigma)$ and
  assume that
  $M(\sigma)$ converges to $\lambda$ when starting in $\rho$.
  Then, for every $f \in \imp(\sigma)$, and $\tup{x}^1,\dots,\tup{x}^m \in D^{ar(f)}$,
  \[
  \sum_{\mathbf{g} \in \mathcal{V}(\sigma)} \rho(\mathbf{g}) f^m(\mathbf{g}(\tup{x}^1,\dots,\tup{x}^m)) \geq 
  \sum_{\mathbf{g} \in \mathcal{V}(\sigma)} \lambda(\mathbf{g}) f^m(\mathbf{g}(\tup{x}^1,\dots,\tup{x}^m)).
  \]
\end{lemma}

\begin{proof}
  By $k$ times applying the $m\rightarrow m$ fractional polymorphism
  $\frac{1}{2} (\chi_{\mathbf{1}} + \sigma)$
  to the left-hand side, we have
  \begin{align*} 
  \sum_{\mathbf{g} \in \mathcal{V}(\sigma)} \rho(\mathbf{g}) f^m(\mathbf{g}(\tup{x}^1,\dots,\tup{x}^m)) & \geq 
  \sum_{\mathbf{g} \in \mathcal{V}(\sigma)} \rho(\mathbf{g})
  \frac{1}{2} \Big(
  f^m(\mathbf{g}(\tup{x}^1,\dots,\tup{x}^m)) \\
  & \qquad + \sum_{\mathbf{h} \in \supp(\sigma)} \sigma(\mathbf{h}) f^m(\mathbf{h} \circ \mathbf{g}(\tup{x}^1,\dots,\tup{x}^m)) \Big) \\
  & = \sum_{\mathbf{g} \in \mathcal{V}(\sigma)} \sum_{\mathbf{g}' \in \mathcal{V}(\sigma)}
  \rho(\mathbf{g}') p(\mathbf{g'},\mathbf{g}) f^m(\mathbf{g}(\tup{x}^1,\dots,\tup{x}^m)) \\
  & \geq \dots \geq \sum_{\mathbf{g} \in \mathcal{V}(\sigma)} \rho^{(k)}(\mathbf{g}) f^m(\mathbf{g}(x^1,\dots,x^m)),
  \end{align*}
  where $\rho^{(k)}(\mathbf{g}) = \sum_{\mathbf{g}' \in \mathcal{V}(\sigma)} \rho(\mathbf{g}') p^{(k)}(\mathbf{g}',\mathbf{g})$.
  By assumption,
  $\rho^{(k)}(\mathbf{g}) \to \lambda(\mathbf{g})$ as $k \to \infty$.
  Since the right-hand side is a linear function in $\rho^{(k)}(\mathbf{g})$,
  the lemma follows by continuity.
\end{proof}


\begin{lemma}\label{lem:ineqs}
  Let $c_1, \dots, c_m \in \Q_{>0}$ and $x_1,\dots,x_m \in \Q$ be such that $\sum_i c_i = 1$, and $x_j \geq \sum_i c_i x_i$ for all $j$.
  Then, $x_j = \sum_i c_i x_i$ for all $j$. 
\end{lemma}

\begin{proof}
  Let $C = \sum_i c_i x_i$. We have $x_j \geq C$ for all $j$.  
  If $x_j > C$ for some $j$,
  then $c_j x_j > c_j C$, so $C = \sum_i c_i x_i > \sum_i c_i C = C$,
  a contradiction.
  So, for all $j$, $x_j \leq C$,
  and hence $x_j = C$.
\end{proof}

\begin{lemma}\label{lem:constant}
  Let $\sigma$ be an $m\to m$ fractional mapping and
  let $C \in \mathcal{R}(\sigma)$.
  Then, for all $f \in \imp(\sigma)$, $\tup{x}^1, \dots, \tup{x}^m \in D^{ar(f)}$,
  and $\mathbf{g} \in C$, 
  \[
  f^m(\mathbf{g}(\tup{x}^1,\dots,\tup{x}^m)) = 
  \sum_{\mathbf{h} \in C} \lambda(\mathbf{h}) f^m(\mathbf{h}(\tup{x}^1,\dots,\tup{x}^m)),
  \] where $\lambda$ is the unique stationary distribution on $M(C)$.
\end{lemma}

\begin{proof}
  For $\mathbf{g} \in C$, let $\chi_{\mathbf{g}}$ be the distribution on $\mathcal{V}$ that
  assigns probability $1$ to $\mathbf{g}$ and 0 to all other mappings in $\mathcal{V}$.
  By Theorem~\ref{thm:markov}(2), $M(\sigma)$ converges to a stationary distribution $\lambda$
  when starting in $\chi_{\mathbf{g}}$.
  By Lemma~\ref{lem:distr},
  \[
  f^m(\mathbf{g}(\tup{x}^1,\dots,\tup{x}^m)) =
  \sum_{\mathbf{h} \in \mathcal{V}} \chi_{\mathbf{g}}(\mathbf{h}) f^m(\mathbf{h}(\tup{x}^1,\dots,\tup{x}^m)) \geq 
  \sum_{\mathbf{h} \in \mathcal{V}} \lambda(\mathbf{h}) f^m(\mathbf{h}(\tup{x}^1,\dots,\tup{x}^m)).
  \]
  
  Note that the chain $M(\sigma)$ stays within the component $C$ when starting in $\chi_{\mathbf{g}}$.
  Therefore,
  $\sum_{\mathbf{h} \in \mathcal{V}} \lambda(\mathbf{h}) f^m(\mathbf{h}(\tup{x}^1,\dots,\tup{x}^m))=
  \sum_{\mathbf{h} \in C} \lambda(\mathbf{h}) f^m(\mathbf{h}(\tup{x}^1,\dots,\tup{x}^m))$,
  and
  $M(C)$ converges to the restriction of $\lambda$ to $C$ when starting in
  the restriction of $\chi_{\mathbf{g}}$ to $C$.
  Hence, by Theorem~\ref{thm:markov}(1), $\lambda(\mathbf{g}) > 0$ for all $\mathbf{g} \in C$.
  It now follows from Lemma~\ref{lem:ineqs} with $c_{\mathbf{g}} = \lambda(\mathbf{g})$ and $x_{\mathbf{g}} = f^m(\mathbf{g}(\tup{x}^1,\dots,\tup{x}^m))$,
  for
  $\mathbf{g} \in C$, that $f^m(\mathbf{g}(\tup{x}^1, \dots, \tup{x}^m)) = \sum_{\mathbf{h} \in C} \lambda(\mathbf{h}) f^m(\mathbf{h}(\tup{x}^1,\dots,\tup{x}^m))$.
\end{proof}

We are now ready to prove Theorem~\ref{thm:master} and Lemma~\ref{lem:mix}.

\begin{proof}[Proof of Theorem~\ref{thm:master}]
  By Theorem~\ref{thm:markov}(2), 
  there exists a stationary distribution $\lambda$ of $M(\sigma)$ such that
  $\sum_{\mathbf{g}'} \sigma(\mathbf{g}') p^{(k)}(\mathbf{g}',\mathbf{g}) \to \lambda(\mathbf{g})$
  as $k \to \infty$,
  for all $\mathbf{g} \in \mathcal{V}(\sigma)$.
  For $C \in \mathcal{R}(\sigma)$, define $w(C) = \sum_{\mathbf{g} \in C} \lambda(\mathbf{g})$.
  By Theorem~\ref{thm:markov}(3), $\lambda(\mathbf{g}) = 0$ for $\mathbf{g} \not\in \mathcal{R}(\sigma)$,
  hence $w$ is a probability distribution on $\mathcal{R}(\sigma)$.

  Let $\rho$ be such that $\sum_{\mathbf{g} \in C} \rho(\mathbf{g}) = w(C)$.
  Arbitrarily pick $f \in \imp(\sigma)$ and $\tup{x}^1, \dots, \tup{x}^m \in D^{ar(f)}$.
  Note that, by Lemma~\ref{lem:distr}, $f \in \imp(\lambda)$.
  Define $\lambda'$ to be the distribution on $C$ given by 
  $\lambda'(\mathbf{g}) = \lambda(\mathbf{g})/w(C)$, for $\mathbf{g} \in C$.
  Then, $\lambda'$ is a stationary distribution on $M(C)$, and by Theorem~\ref{thm:markov}(1),
  it is unique.
  Therefore, by Lemma~\ref{lem:constant}, we have
  \begin{align*}
  \sum_{\mathbf{g} \in C} \rho(\mathbf{g}) f^m(\mathbf{g}(\tup{x}^1, \dots, \tup{x}^m)) &=
  \sum_{\mathbf{g} \in C} \rho(\mathbf{g}) \sum_{\mathbf{h} \in C} \lambda'(\mathbf{h}) f^m(\mathbf{h}(\tup{x}^1, \dots, \tup{x}^m))\\
  &= w(C) \sum_{\mathbf{h} \in C} \lambda'(\mathbf{h}) f^m(\mathbf{h}(\tup{x}^1, \dots, \tup{x}^m))\\
  &= \sum_{\mathbf{h} \in C} \lambda(\mathbf{h}) f^m(\mathbf{h}(\tup{x}^1, \dots, \tup{x}^m)).
  \end{align*}
  As this holds for every $C \in \mathcal{R}(\sigma)$, it follows that $f \in \imp(\rho)$.
\end{proof}

\subsection{Proof of Lemma~\ref{lem:mix}}
\label{sec:proofs2}

\begin{proof}[Proof of Lemma~\ref{lem:mix}]
  Let $C \in \mathcal{R}(\sigmaour)$ be the component containing $\mathbf{g}$,
  and for $i = 1, 2$, let $\Omega_i = \{ \mathbf{h} \in \supp(\rhoour) \mid \mathbf{h}(a_1,a_2) = (a_i,a_i) \}$.
  \begin{align}
    f^2((a_1,\tup{y}^1), (a_2,\tup{y}^2)) 
    & \geq 
    \sum_{\mathbf{h} \in \supp(\rhoour)} \rhoour(\mathbf{h}) f^2(\mathbf{h}((a_1,\tup{y}^1), (a_2,\tup{y}^2)))
    \label{eq:1} \\
  & =
  \sum_{\mathbf{h} \in \Omega_1} \rhoour(\mathbf{h})
  f^2(\mathbf{h}((a_1,\tup{y}^1),(a_1,\tup{y}^2))) \nonumber \\ & \qquad +
  \sum_{\mathbf{h} \in \Omega_2} \rhoour(\mathbf{h})
  f^2(\mathbf{h}((a_2,\tup{y}^1),(a_2,\tup{y}^2))) \label{eq:2} \\
  & =
  \frac{1}{2} 
  f^2((a_1,\tup{y}^1),(a_1,\tup{y}^2)) +
  \frac{1}{2} 
  f^2((a_2,\tup{y}^1),(a_2,\tup{y}^2))\label{eq:3} \\
  & =
  \frac{1}{2}
  f^2((a_1,\tup{y}^1),(a_2,\tup{y}^2)) +
  \frac{1}{2}
  f^2((a_1,\tup{y}^2),(a_2,\tup{y}^1)), \label{eq:4}
  \end{align}
  where
  (\ref{eq:1}) follows by applying $\rhoour$ and
  (\ref{eq:2}) follows from $\rhoour$ being idempotent and submodular on $\{a_1,a_2\}$.
  To obtain (\ref{eq:3}),
  note that $\mathbf{h} \circ \mathbf{g} \in C$, so by the first part of Lemma~\ref{lem:constant},
  $f^2(\mathbf{h} \circ \mathbf{g}((a_i,\tup{x}^1),(a_i,\tup{x}^2))) =
  f^2(\mathbf{g}((a_i,\tup{x}^1),(a_i,\tup{x}^2))) = f^2((a_i,\tup{y}^1),(a_i,\tup{y}^2))$ for all $\mathbf{h} \in \Omega_i$ and $i = 1, 2$.
  Finally,
  (\ref{eq:4}) follows by rearranging the terms.

  This shows the inequality $f^2((a_1,\tup{y}^1),(a_2,\tup{y}^2)) \geq f^2((a_1,\tup{y}^2),(a_2,\tup{y}^1))$.
  The reverse inequality follows analogously.
\end{proof}

\section{Symmetric fractional polymorphisms of all arities} 
\label{app:binary}

An important step in the proof of Theorem~\ref{thm:blp} is showing that a binary
symmetric fractional polymorphism ``generates'' symmetric fractional
polymorphisms of all higher arities. 
This was proved in \cite{Kolmogorov13:icalp}.
In this section, we demonstrate the power
of the Markov chain machinery set up in Section~\ref{sec:proofs} by giving an
alternative proof of this theorem.
The proof idea is the same as that of \cite{Kolmogorov13:icalp},
but the proof is substantially shortened.

\begin{theorem}[\cite{Kolmogorov13:icalp}] \label{thm:bintoarb}
  Suppose $\Gamma$ is a constraint language with a symmetric fractional
  polymorphism of arity $2$.
  Then $\Gamma$ has symmetric fractional polymorphisms of all arities.
\end{theorem}

\begin{proof} 
It suffices to prove that if $\Gamma$ has a symmetric fractional polymorphism
of arity $m-1 \geq 2$, then it has one of arity $m$.
Let $\omega$ be an $(m-1)$-ary symmetric fractional polymorphism of $\Gamma$.
For $1 \leq k \leq m$,
let $\delta_k \in \opm{D}{m \to m-1}$ denote the mapping obtained
by omitting the $k$th operation from the identity mapping in $\opm{D}{m \to m}$.
Define
\[
\sigma := \sum_{h \in \supp(\omega)} \omega(h) \chi_{(h \circ \delta_1, \dots, h \circ \delta_m)}.
\]

Then, $\sigma$ is a fractional polymorphism of $\Gamma$.
Let $\rho$ be a fractional polymorphism of $\Gamma$ of arity $m \to m$
as given by Theorem~\ref{thm:master} applied to $\sigma$,
and let $\mathbf{p}$ be any symmetric and permuting mapping of arity $m \to m$.
For example, let $\mathbf{p}$ be a mapping that orders its $m$ inputs according to some fixed total order on $D$.
We claim that $\rho' = \rho \circ \mathbf{p}$ is a fractional
polymorphism of $\Gamma$, from which the theorem follows as $\rho'$ is clearly symmetric.

Let $f\in \Gamma$ and $\tup{x}^1,\dots,\tup{x}^m \in D^{ar(f)}$.
It suffices to show that for every $\mathbf{g} \in \supp(\rho)$,
$f^m(\mathbf{g}(\tup{x}^1,\dots,\tup{x}^m)) = f^m(\mathbf{g} \circ \mathbf{p}(\tup{x}^1,\dots,\tup{x}^m))$.
We do this by showing that for any $1 \leq i \leq ar(f)$ and $1 \leq j_1, j_2 \leq m$,
interchanging $x^{j_1}_i$ and $x^{j_2}_i$ does not alter the value of
$f^m(\mathbf{g}(\tup{x}^1,\dots,\tup{x}^m))$.
The result then follows by repeatedly interchanging such pairs of elements in $(\tup{x}^1,\dots,\tup{x}^m)$ to obtain $\mathbf{p}(\tup{x}^1,\dots,\tup{x}^m)$.

  For $1 \leq k \leq m$, let $\pi_k \in \opm{D}{m}$ denote the projection on the $k$th component.
Since $m \geq 3$, we can pick $k \in \{1,\dots,m\} \setminus
 \{j_1, j_2\}$.
Let $\mathbf{h} \in \supp(\sigma)$ and let $\tau$ be a permutation on $\{1,\dots,m\}$ that interchanges $j_1$ and $j_2$.
By definition of $\sigma$,
\begin{equation}\label{eq:commute}
\pi_k \circ \mathbf{h}(x_1, \dots, x_m) = \pi_k \circ \mathbf{h}(x_{\tau(1)},\dots,x_{\tau(m)}),
\end{equation}
for $x_1, \dots, x_m \in D$.
Furthermore, this identity is seen to hold for any
$\mathbf{h} = \mathbf{h}_{\ell} \circ \dots \circ \mathbf{h}_1 \in \mathcal{V}(\sigma)$ by induction over $\ell$.

Let $C \in \cal{R}(\sigma)$ be the component containing $\mathbf{g}$ and let $\lambda$ be
the unique stationary distribution on $M(C)$. 
Then we have
  \begin{align}
  \sum_{\mathbf{h} \in C} \lambda(\mathbf{h}) f^{m-1}(\delta_k \circ \mathbf{h}(\tup{x}^1,\dots,\tup{x}^m))
  &\geq
  \sum_{\mathbf{h} \in C} \lambda(\mathbf{h}) \sum_{h \in \supp(\omega)} \omega(h) f(h \circ \delta_k \circ \mathbf{h}(\tup{x}^1,\dots,\tup{x}^m) \notag \\
  &= 
  \sum_{\mathbf{h} \in C} \lambda(\mathbf{h}) \sum_{\mathbf{h}' \in \supp(\sigma)} \sigma(\mathbf{h}') f(\pi_k \circ \mathbf{h}' \circ \mathbf{h}(\tup{x}^1,\dots,\tup{x}^m)) \notag \\
  &=
  \sum_{\mathbf{h} \in C} \lambda(\mathbf{h}) \cdot 2 \sum_{\mathbf{h}' \in C} p(\mathbf{h},\mathbf{h}') f(\pi_k \circ \mathbf{h}'(\tup{x}^1,\dots,\tup{x}^m)) \notag \\
  &\qquad - \sum_{\mathbf{h} \in C} \lambda(\mathbf{h}) f(\pi_k \circ \mathbf{h}(\tup{x}^1,\dots,\tup{x}^m)), \notag \\
  &=
  \sum_{\mathbf{h} \in C} \lambda(\mathbf{h}) f(\pi_k \circ \mathbf{h}(\tup{x}^1,\dots,\tup{x}^m)),
  \label{eq:9000b}
\end{align}
where the inequality follows from applying (\ref{ineq:fpol}) to $\omega$, 
the first equality follows from the definition of $\sigma$, the
second equality follows from the definition of the transition probabilities for $M(C)$:
\begin{align*}
\sum_{\mathbf{h}' \in C} p(\mathbf{h},\mathbf{h}') f(\pi_k \circ \mathbf{h}'(\tup{x}^1,\dots,\tup{x}^m))
&= \frac{1}{2} \sum_{\mathbf{h}' \in \supp(\sigma)} \sigma(\mathbf{h}') f(\pi_k \circ \mathbf{h}' \circ \mathbf{h}(\tup{x}^1,\dots,\tup{x}^m))\\
& +
\frac{1}{2} f(\pi_k \circ \mathbf{h}(\tup{x}^1,\dots,\tup{x}^m)),
\end{align*}
and the third equality follows by interchanging the order of summation in the first part and then
using the fact that $\lambda$ is the stationary distribution of $M(C)$.
 By (\ref{eq:9000b}) and Lemma~\ref{lem:ineqs} with $c_{k} = \frac{1}{m}$ and $x_{k} = - \sum_{\mathbf{h} \in C} \lambda(\mathbf{h}) f(\pi_k \circ \mathbf{h}(\tup{x}^1,\dots,\tup{x}^m))$,
 we have $
  \sum_{\mathbf{h} \in C} \lambda(\mathbf{h}) f^m(\mathbf{h}(\tup{x}^1,\dots,\tup{x}^m)) =
  \sum_{\mathbf{h} \in C} \lambda(\mathbf{h}) f(\pi_k \circ \mathbf{h}(\tup{x}^1,\dots,\tup{x}^m)
  $,
  so
by Lemma~\ref{lem:constant}, it follows that
\begin{equation}\label{eq:9000}
f^m(\mathbf{g}(\tup{x}^1,\dots,\tup{x}^m)) 
= \sum_{\mathbf{h} \in C} \lambda(\mathbf{h}) f(\pi_k \circ \mathbf{h}(\tup{x}^1,\dots,\tup{x}^m)).
\end{equation}
By (\ref{eq:commute}), interchanging $x^{j_1}_i$ and $x^{j_2}_i$ does not alter the value of the
right-hand side of (\ref{eq:9000}) and hence it does not alter the value of
$f^m(\mathbf{g}(\tup{x}^1,\dots,\tup{x}^m))$.
The result follows.
\end{proof}

\section{Conclusions}

In this work we have completely answered the question of which finite-valued
constraint languages on finite domains are solvable exactly in polynomial time.
In particular, we have characterised the tractable constraint languages as those that admit
a binary symmetric fractional polymorphism.
We have also shown tractability
to be a polynomial-time checkable condition, assuming that the constraint language is a core. 
By previous results, this implies that all tractable constraint languages
are solvable by the basic linear programming relaxation. Thus, we have
demonstrated that the basic linear programming (BLP) relaxation suffices for exact
solvability of finite-valued constraint languages and that, in this context,
semidefinite programming relaxations do not add any power.

\section{Acknowledgments}

The authors wish to thank Hubie Chen for useful comments on an earlier draft
of this paper, Vladimir Kolmogorov for pointing out that Theorem~\ref{thm:main}
holds without requiring the constraint language $\Gamma$ to be a core, and the
anonymous referees for their thorough work.


\newcommand{\noopsort}[1]{}

\appendix

\section{Infinite constraint languages}
\label{sec:infinite}

The main result of this article, Theorem~\ref{thm:dichotomy}, establishes a
dichotomy for finite-valued constraint languages of \emph{finite} size. The
finiteness is important when passing from the primal to the dual linear
programme, and thus could be considered an artefact of our proof techniques.
However, our algorithm, the BLP, only depends on the instance and not in some
exponential way on the constraint language. We are therefore able to extend our
results to finite-valued constraint languages of \emph{infinite} size; that is,
the setting when the cost functions are still represented extensionally.

To state the dichotomy for infinite constraint languages,
we need to allow the fractional
polymorphisms to take on \emph{real} values. Hence for the rest of this
section, an $m$-ary fractional operation is a function $\omega: \opm{D}{m}
\rightarrow \R_{\geq 0}$, $\|\omega\|_1=1$. Fractional polymorphisms are defined
by inequality~(\ref{ineq:fpol}), using real-valued fractional operations. Note
however that the constraint languages, although infinite, still consist of
rational-valued cost functions only. 

\begin{theorem}\label{thm:infinite-dichotomy}
Let $D$ be an arbitrary finite set, let $\Gamma$ be a (possibly infinite)
constraint language defined on $D$, and
let $\Gamma'$ be a core of $\Gamma$.
\begin{itemize}
\item Either $\Gamma$ has a binary symmetric real-valued fractional polymorphism and BLP
solves VCSP$(\Gamma)$;
\item or (MC) holds for $\Gamma'_c$ and VCSP$(\Gamma)$ is NP-hard.
\end{itemize}
\end{theorem}

It follows from~\cite{tz12:focs,Kolmogorov13:icalp,ktz15:sicomp} that for a
(possibly infinite) constraint language $\Gamma$ with a binary symmetric
real-valued fractional polymorphism, $\Gamma$ is not only tractable but also
\emph{globally tractable}.
Conversely, we need to show that if $\Gamma$ does not have a binary
symmetric fractional polymorphism, then the same holds for some finite subset of
$\Gamma$. We can then apply Theorem~\ref{thm:main} to conclude that $\Gamma$ is
NP-hard. This direction is a consequence of the following result, when $\Omega$
is taken as the set of symmetric $m$-ary operations on $D$. A similar result for
countably infinite constraint languages is proved in~\cite{ktz15:sicomp}.

\begin{lemma}\label{lem:compact}
 Let $\Gamma$ be a (possibly infinite) constraint language. Let $\Omega\subseteq
 \opm{D}{m}$ and assume that every finite subset $\Gamma'\subseteq\Gamma$ has a
 fractional polymorphism with support in $\Omega$. Then $\Gamma$ has a fractional
 polymorphism with support in $\Omega$.
\end{lemma}

\begin{proof}
 Note that $|\Omega|$ is finite and let $n = |\Omega|$.
 Let $Y$ be the set of fractional operations
 $\omega:\Omega\rightarrow\R_{\geq 0}$, $\|\omega\|_1 = 1$. 
 Then $Y$ is a compact set in $\R^n$.
 Assume, for the sake of contradiction, that $\Gamma$ does not have a fractional polymorphism
 with support in $\Omega$. 
 Then, for every $y \in Y$, there is
 some $f_y \in \Gamma$ and $\tup{x}^1,\dots,\tup{x}^k \in D^{ar(f_y)}$ such
 that
 \[
   \sum_{g \in \Omega} y(g) f_y(g(\tup{x}^1,\dots,\tup{x}^k)) > 
   f^m_y(\tup{x}^1,\dots,\tup{x}^k).
 \]
 Furthermore, this inequality holds in an open neighbourhood $U_y \subseteq Y$
 of $y$, so $\{U_y\}_{y \in Y}$ is an open cover of $Y$. Since every open cover
 of a compact set has a finite subcover, this provides us with a finite subset
 of $\Gamma$ that does not have a fractional polymorphism with support in
 $\Omega$. This is a contradiction, hence $\Gamma$ must have a fractional
 polymorphism with support in $\Omega$. 
\end{proof}

The proof of Lemma~\ref{lem:compact} relies on real-valued fractional
polymorphisms,
and the obvious question to ask is then whether real values
are necessary for Theorem~\ref{thm:infinite-dichotomy} to hold, 
or whether it is an artefact of our proof techniques.
Perhaps unsurprisingly, we can demonstrate that real-valued fractional
polymorphisms are necessary in \emph{some} cases.
The following construction is based on a language from~\cite{hkp14:sicomp}, where it was
used for a different result; we will use the same notation as
in~\cite{hkp14:sicomp}.

Let $D=\{-1,0,1\}$ and fix the partial order $-1>0<1$ on $D$. For $a\in\{-1,1\}$, define binary operations
$\vee_a$ and $\wedge_0$ as follows:
\[
1 \vee_a -1 = -1 \vee_a 1 = a \mbox{ and } x\vee_a y=\max(x,y) \mbox{ wrt the above order if } \{x,y\}\neq\{-1,1\};
\]
\[
1 \wedge_0 -1 = -1 \wedge_0 1 = 0 \mbox{ and } x\wedge_0 y=\min(x,y) \mbox{ wrt the above order if } \{x,y\}\neq\{-1,1\}.
\]

Let $\alpha \in (0,1]$ be an arbitrary real constant, and define the fractional operation
$\omega$ as follows: $\omega(\wedge_0)=1/2$, $\omega(\vee_0)=\alpha/2$, and
$\omega(\vee_1)=(1-\alpha)/2$.
A cost function is called \emph{$\alpha$-bisubmodular} if it admits the fractional polymorphism $\omega$.

For an arbitrary rational $\alpha\in(0,1]$, write $\alpha=p/q$ with $p,q\geq 1$, $p$ and $q$ coprime.
Define the unary cost
functions $e, {u_\alpha, v_\alpha:D\to\range}$ and the binary cost function
$f:D^2\to\range$ as follows:

\begin{center}
\begin{tabular}{c|c|c|c}
 & $-1$ & $0$ & $1$ \\ \hline
$e$ & $1$ & $0$ & $1$ \\
$u_\alpha$ & $p+q$ & $q$ & $0$ \\
$v_\alpha$ & $0$ & $p$ & $p+q$ \\
\end{tabular}
\qquad
\qquad
\qquad
\begin{tabular}{r|c|c|c}
$f$ & $-1$ & $0$ & $1$ \\ \hline
$-1$ & $3$ & $2$ & $1$ \\
$0$  & $2$ & $0$ & $0$ \\
$1$  & $1$ & $0$ & $0$ \\
\end{tabular}
\end{center}
Note that $u_\alpha$ and $v_\alpha$ are uniquely defined given $\alpha$.

\begin{proposition}
Fix an arbitrary irrational value $x \in (0,1)$ and define
\[\Gamma_x := 
\{ v_\alpha \mid \alpha\in \Q \cap (0,x) \}\cup
\{ u_\alpha \mid \alpha\in \Q \cap (x,1] \}\cup
\{ e, f \}.
\]
\begin{enumerate}
\item
$\Gamma_x$ is $x$-bisubmodular and BLP solves VCSP$(\Gamma_x)$, but
\item
$\Gamma_x$ does not admit any rational-valued binary symmetric fractional
polymorphism.
\end{enumerate}
\end{proposition}

\begin{proof}
We first show part (1).
It follows from the definition that 
unary function $u$ is $x$-bisubmodular if, and only if,
\begin{equation}\label{unary}
(1+x) \cdot u(0)\leq u(-1)+x \cdot u(1).
\end{equation}

For the cost function $e$, condition (\ref{unary}) becomes $(1+x) \cdot 0 \leq 1 + x \cdot 1$, 
so $e$ is $x$-bisubmodular.
For the cost function $u_\alpha$, since $x < \alpha = p/q$, we have
$(1+x) q < p+q$, so (\ref{unary}) holds and $u_\alpha$ is $x$-bisubmodular.
Similarly, one shows that $v_\alpha$ is $x$-bisubmodular for $x > \alpha$.

It remains to show that $f$ is $x$-bisubmodular. 
By an alternative characterisation~\cite[Proposition 2]{hkp14:sicomp},
$f$ is $x$-bisubmodular if and only if
(i) the unary cost functions obtained from $f$ by fixing one argument are
$x$-bisubmodular, and (ii) $f$ is submodular in every orthant;
this means that for
every $\tup{c}\in\{-1,1\}^2$, the fractional polymorphism
inequality~(\ref{ineq:fpol}) holds for $x$-bisubmodularity for all
$\tup{a}^1,\tup{a}^2\in D^2$ with $\tup{a}^1,\tup{a}^2\leq \tup{c}$ (here we
used the componentwise order on $D$).

First we verify that the unary cost functions
$f(-1,x)$, $f(0,x)$, and $f(1,x)$ are $x$-bisubmodular.
The inequality (\ref{unary}) becomes $(1+x) \cdot 2 \leq 3+x$,
$(1+x) \cdot 0 \leq 2+0 \cdot x$, and $(1+x) \cdot 0 \leq 1+x$, respectively.
Since $x \in (0,1)$, all three inequalities hold, so all three cost functions are $x$-bisubmodular.
By symmetry, $f(x,-1)$, $f(x,0)$, and $f(x,1)$ are also $x$-bisubmodular.

Next, we verify that $f$ is submodular in every orthant:
\begin{itemize}
\item $f$ is constant $0$ and hence trivially submodular in the orthant $(1,1)$.
\item
In the orthant $(-1,-1)$, the only
nontrivial case to verify is $\tup{a}^1=(0,-1)$ and $\tup{a}^2=(-1,0)$. We have, after
multiplying by $2$, $f(0,-1)+f(-1,0)=2 + 2 \geq
1 \cdot f(0,0)+
x \cdot f(-1,-1) +
(1-x) \cdot f_\alpha(-1,-1) = f(-1,-1) = 3$,
which holds true.
Hence, $f$ is submodular in the orthant $(-1, -1)$.
\item
Finally, the two cases $\tup{c} = (1,-1)$ and $\tup{c} = (-1,1)$ are symmetric.
In the orthant $(1,-1)$, the only nontrivial case to verify is $\tup{a}^1=(0,-1)$ and
$\tup{a}^2=(1,0)$. Here, we have
$f(0,-1)+f_\alpha(1,0)=2+0 \geq f(0,0)+x f(1,-1)+(1-x) \cdot f(1,-1)=f_\alpha(1,-1)=1$, 
which holds true.
Hence, $f$ is submodular in the orthants $(1,-1)$ and $(-1,1)$.
\end{itemize}
We conclude that $\Gamma_x$ is $x$-bisubmodular, and hence solved by the BLP relaxation.

We now show part (2).
Let $\omega$ be an arbitrary binary symmetric fractional polymorphism of $\Gamma_x$. For
$a\in\{-1,0,1\}$, define $w_a=\sum_{g\in\opm{D}{2}\:|\:g(-1,1)=a}
\omega(g)$. Clearly, $0\leq w_a\leq 1$ and
$w_{-1}+w_0+w_1=1$. It suffices to show that at least one of the
$w_a$ is irrational, $a\in\{-1,0,1\}$, which implies the existence of a
binary operation $g$ with $\omega(g)\not\in\Q$.

Let $\alpha=p/q$ with $\alpha<x$. Applying the fractional polymorphism
inequality~(\ref{ineq:fpol}) to $v_\alpha\in\Gamma_x$, we have
$(p+q)/2 = (v_\alpha(-1)+v_\alpha(1))/2  \geq
w_0v_\alpha(0)+w_1v_\alpha(1)+w_{-1}v_\alpha(-1) =
w_0p+w_1(p+q)+w_{-1}0$, which is equivalent to $w_0\leq
(1+1/\alpha)(1/2-w_{1})$. 
Since this inequality holds for all rational $\alpha < x$,
we have, in the limit as $\alpha \to x$ from below,
\begin{equation}\label{eq:w1}
w_0\leq (1+1/x)(1/2-w_{1}).
\end{equation}

A similar argument for the cost function $u_\alpha \in \Gamma_x$, for $\alpha > x$, 
leads to the inequality $w_0 \leq (1+\alpha)(1/2-w_{-1})$
and, in the limit as $\alpha \to x$ from above,
\begin{equation}\label{eq:w2}
w_0 \leq  (1+x)(1/2-w_{-1}).
\end{equation}

Add $x$ times the inequality (\ref{eq:w1}) to the inequality (\ref{eq:w2})
to obtain $(1+x)w_0 \leq (1+x)(1-w_{1}-w_{-1})$.
Since $w_{-1}+w_0+w_1 = 1$, this inequality must hold with equality, 
and hence the
inequalities (\ref{eq:w1}) and (\ref{eq:w2}) can be replaced by the equalities
$w_0 = (1+1/x)(1/2-w_1)$ and $w_0 = (1+x)(1/2-w_{-1})$.
Since $x$ is irrational, it follows that either 
$w_0 = 0$ and $w_{-1} = w_{1} = 1/2$,
or at least one of
$w_{-1}$, $w_0$, and $w_1$ is irrational.
We demonstrate that the latter holds by showing that $w_{-1} < 1/2$.
Applying the fractional polymorphism inequality~(\ref{ineq:fpol}) to
$f\in\Gamma_x$,
we have
$1=(1+1)/2=(f(-1,1)+f(1,-1))/2\geq w_{-1}f(-1,-1)+w_0 f(0,0) + w_1 f(1,1) 
= w_{-1} \cdot 3$, which gives $w_{-1}\leq 1/3 < 1/2$, 
and the claim follows.
\end{proof}

\end{document}